\documentstyle[12pt]{article} 
\begin{document}
\oddsidemargin .4cm
\begin{titlepage}
\title{
\hfill {\normalsize IFT/1/95}\\
\hfill{\normalsize hep-ph/9602245}\\
\hfill{\normalsize revised version Dec. 1995}\\
\vspace{1cm}
 Improved Evaluation of the Next-Next-To-Leading Order
 QCD Corrections to the $e^{+}e^{-}$ Annihilation Into
 Hadrons
 }
\author{
 Piotr A. R\c{a}czka\thanks{E-mail: praczka@fuw.edu.pl}
 \hspace{.3cm}
 and  Andrzej Szymacha\\ 
 {\small Institute of Theoretical Physics}\\
 {\small  Department of Physics, Warsaw University}\\
 {\small ul.\ Ho\.{z}a 69, PL-00-681 Warsaw, Poland.}
 }
\date{\quad}
\maketitle
\begin{abstract}
\noindent 
The next-next-to-leading order QCD corrections to the
$e^{+}e^{-}$ annihilation into hadrons are considered. The
stability of the predictions with respect to change of the
renormalization scheme is discussed in detail for the case of
five, four and three active quark flavors. The analysis is based
on the recently proposed condition for selecting renormalization
schemes according to the degree of cancellation that they
introduce in the expression for the scheme invariant combination
of the expansion coefficients. It is demonstrated that the scheme
dependence ambiguity in the predictions obtained with the
conventional expansion is substantial, particularly at lower
energies. It is shown however, that the stability of the
predictions is greatly improved when QCD corrections are
evaluated in a more precise way, by utilizing the contour
integral representation and calculating numerically the contour
integral.
\end{abstract}
\begin{center}
\end{center}
\thispagestyle{empty}
\end{titlepage}
\setcounter{page}{1} 
\newpage
\section{Introduction}

In a series of recent papers \cite{racz92}-\cite{racz95b} a
method has been presented for a systematic analysis of the
renormalization scheme (RS) ambiguities in the
next-next-to-leading (NNLO) perturbative QCD predictions. It was
emphasized in \cite{racz92,racz93,racz95a} that besides giving
predictions in some preferred renormalization scheme one should
also investigate the stability of the predictions when the
parameters determining the scheme are changed in some acceptable
range. The method discussed in \cite{racz92,racz93,racz95a}
involves a specific condition that allows one to eliminate from
the analysis the renormalization schemes that give rise to
unnaturally large expansion coefficients. The condition on the
acceptable schemes is based on the existence in NNLO of the RS
invariant combination of the expansion coefficients, which is
characteristic for the considered physical quantity. The method
of \cite{racz92,racz93,racz95a} has been applied to the QCD
corrections to the Bjorken sum rule for the polarized structure
functions \cite{racz95a} and to the QCD corrections to the total
hadronic width of the tau lepton \cite{racz92,racz93,racz94}.

In this note we apply this method to the QCD correction to
$R_{e^{+}e^{-}}$ ratio:
\begin{equation} 
R_{e^{+}e^{-}}=\frac{
\sigma_{tot}(e^{+}e^{-} \rightarrow \mbox{hadrons})
}{
\sigma_{tot}(e^{+}e^{-} \rightarrow \mu^{+}\mu^{-})
}.
\end{equation} 
which received considerable attention in recent years
\cite{gori88}-\cite{jone95}. We show that a straightforward
application of the condition proposed in
\cite{racz92,racz93,racz95a} to the conventional perturbative
expression for the QCD effects in the $R_{e^{+}e^{-}}$ ratio
exhibits a rather strong RS dependence, even at high energies.
Looking for improvement and motivated by the analysis of the
corrections to the tau decay
\cite{racz92,racz93,pivo92b,ledi92,racz94}, we calculate the QCD
correction to the $R_{e^{+}e^{-}}$ ratio by using the contour
integral representation \cite{radu82,radu86} and
evaluating the contour integral numerically. In this way we
resumm to all orders some of the so called $\pi^{2}$ corrections,
which appear as a result of analytic continuation of the
expression for the hadronic vacuum polarization function from
spacelike to timelike momenta \cite{moor77,pivo92a}. Such
corrections constitute a dominant contribution in the NNLO. 
Using the improved expression we perform similar analysis as in
the case of the conventional expansion. We find that the
predictions obtained by numerical evaluation of the contour
integral show extremely good stability with respect to change of
the RS.

The results reported here have been announced in \cite{racz95a} and
briefly described in \cite{racz95b}.

\section{$\delta_{e^{+}e^{-}}$ and the problem of renormalization
scheme ambiguity} 

Away from the thresholds, neglecting the effects of the quark
masses and the electroweak corrections, the formula for
$R_{e^{+}e^{-}}$ may be written in the form:
\begin{equation}
R_{e^{+}e^{-}}(s)= 3 \sum_{f} Q^{2}_{f} [ 1 + \delta_{e^{+}e^{-}}(s)],
\label{Ree}
\end{equation}
where $Q_{f}$ denotes the electric charge of the quark with the
flavor $f$ and $\delta_{e^{+}e^{-}}$ is the QCD correction. The
renormalization group improved NNLO expression for
$\delta_{e^{+}e^{-}}$ has the form:
\begin{equation}
\delta^{(2)}_{e^{+}e^{-}}(s) = a(s)\,
[ 1 + r_{1} a(s) + r_{2} a^{2}(s)],
\label{delta}
\end{equation}
where $a(\mu^{2})=g^{2}(\mu^{2})/(4\pi^{2})$ is the coupling
constant, satisfying the renormalization group equation:
\begin{equation}
\mu \frac{da}{d\mu} = - b\,a^{2}\,(1 + c_{1}a + c_{2}a^{2}\,).
\label{rge}
\end{equation}
The perturbative result for $\delta^{(2)}_{e^{+}e^{-}}$ is
usually expressed in the Modified Minimal Subtraction
($\overline{\mbox{MS}}$) renormalization scheme,
i.e. using the $\overline{\mbox{MS}}$ renormalization convention
\cite{bard78} 
with $\mu^{2}=s$. In the $\overline{\mbox{MS}}$ scheme we have
\cite{chet79,surg91,gori91}:
\begin{eqnarray}
r^{\overline{MS}}_{1}&=&1.985707-0.115295\,n_{f},\\
r^{\overline{MS}}_{2}&=&18.242692-4.215847\,n_{f} +
0.086207\,n^{2}_{f}+\nonumber\\
 & & +\,r_{2}^{sing}-(b\pi/2)^{2}/3,
\end{eqnarray}
where the $r_{2}^{sing}$ term in $r_{2}^{\overline{MS}}$
represents the so called flavor singlet contribution:
\begin{equation}
r_{2}^{sing}=
\frac{(\sum_{f} Q_{f})^2}{\sum_{f} Q_{f}^{2}} 
\left(\frac{55}{216} - \frac{5}{9} \zeta_{3}\right),
\end{equation}
which arises from the light-by-light scattering type of diagrams
($\zeta_{3} = 1.202056903$). (It should be noted that the first
calculation of the NNLO correction \cite{gori88} was erroneous.
The corrected result was published in \cite{gori91}. An
independent evaluation was reported in \cite{surg91}.) For the
coefficients in the renormalization group equation we have
$b=(33-2n_{f})/6$, $c_{1}=(153-19n_{f})/(66-4n_{f})$ and
\cite{tara80}:
\begin{equation}
c_{2}^{\overline{MS}}=
\frac{77139-15099\,n_{f}+325\,n_{f}^{2}}{288(33-2\,n_{f})}.
\end{equation}
 For convenience we collect in Table~1 the numerical values of
the expansion coefficients for various values of $n_{f}$.
\begin{table}
\begin{center}
\begin{tabular}{||c|c|c|c|c|c|c||}
\hline
$n_{f}$ 
    &$r_{1}^{\overline{MS}}$
              &$r_{2}^{\overline{MS}}$
                       &$r_{2}^{sing}$
                           &$c_{1}$
                               &$c_{2}^{\overline{MS}}$
                                      &$\rho_{2}^{R}$  \\
\hline
2& 1.75512&  -9.14055& -0.08264& 1.98276&  5.77598&  -9.92498\\
\hline
3& 1.63982& -10.28394&  0.00000& 1.77778&  4.47106& -11.41713\\
\hline
4& 1.52453& -11.68560& -0.16527& 1.54000&  3.04764& -13.30991\\
\hline
5& 1.40923& -12.80463& -0.03756& 1.26087&  1.47479& -15.09262\\
\hline
6& 1.29394& -14.27207& -0.24791& 0.92857& -0.29018& -17.43803\\
\hline
\end{tabular}   
\end{center}
\caption{ Numerical values of the expansion coefficients $r_{i}$
 for $\delta^{(2)}_{e^{+}e^{-}}$, obtained with the
 $\overline{\mbox{MS}}$ renormalization convention and
 $\mu^{2}=s$, for various numbers of quark flavors. The magnitude
 of the flavor singlet contribution $r_{2}^{sing}$ is separately
 indicated. The values of the RS invariant $\rho_{2}^{R}$ are
 calculated according to Eq.~(\protect\ref{rho2}). The numerical values
 of the coefficients $c_{i}$ in the renormalization group
 equation are included for completeness.}
\end{table}

Besides the $\overline{\mbox{MS}}$ scheme other choices of the RS
are of course possible, such as for example the momentum
subtraction schemes \cite{celm79b}. A change
in the RS modifies the values of the expansion coefficients ---
the relevant formulas have been collected for example in
\cite{racz92}. (The coefficients $b$ and $c_{1}$ are RS
independent in the class of mass and gauge independent schemes.)
The change in the expansion coefficients compensates for the
finite renormalization of the coupling constant. Of course, in
the given order of perturbation expansion this compensation may
be only approximate, so that there is some numerical difference
in the perturbative predictions in various schemes. This
difference is formally of higher order in the coupling --- it is
$O(a^{4})$ for the NNLO expression --- but numerically the
difference may be significant for comparison of theoretical
predictions with the experimental data. There has been a lively
discussion how to avoid this problem, both in the general case
\cite{grun80}-\cite{dhar83} (for a summary of early contributions
see \cite{duke85})  and in the particular case of
$\delta_{e^{+}e^{-}}$ \cite{maxw88}-\cite{surg93}. (Some of the
early papers \cite{maxw88}-\cite{dhar90} contain discussion of
$\delta^{(2)}_{e^{+}e^{-}}$ with the erroneous value of the NNLO
correction reported in \cite{gori88}. Much of the initial
interest in the RS dependence of $\delta^{(2)}_{e^{+}e^{-}}$ came
from the fact that this erroneous correction was very large.) It
seems that one
of the most interesting propositions is to choose the scheme
according to the so called Principle of Minimal Sensitivity (PMS)
\cite{stev81}.

However, as was emphasized in \cite{racz92,racz93,racz95a}, besides
calculating the predictions in some preferred renormalization
scheme, it is also important to investigate the stability of the
predictions with respect to reasonable variations in the scheme
parameters. By calculating the variation in the predictions over
some set of {\em a~priori} acceptable schemes one obtains a
quantitative estimate of reliability of the optimized
predictions. A systematic method for analyzing the stability of 
predictions with respect to change of the renormalization scheme
has been presented in \cite{racz92,racz93,racz95a}. This method is
based on the existence of the RS invariant combination of the
expansion coefficients \cite{grun80,stev81,dhar83}:
\begin{equation}
\rho_{2}=c_{2}+r_{2}-c_{1}r_{1}-r_{1}^{2}, 
\label{rho2}
\end{equation}
which appears to be a natural RS independent characterization of
the magnitude of the NNLO correction. (We adopt here the
definition of the RS invariant used in \cite{grun80,dhar83},
which differs by a constant from the definition of Stevenson
\cite{stev81}: $\rho_{2}^{Stev}=\rho_{2}-c_{1}^{2}/4$. The
arguments in favor of Eq.~(\ref{rho2}) have been given in
\cite{racz95a}.) The numerical values of this invariant in the
case of $\delta_{e^{+}e^{-}}$, for different values of $n_{f}$,
are collected in Table~1.

   The $\rho_{2}$ invariant may be used to eliminate from the
analysis the unnatural
renormalization schemes. This is done by
introducing a function $\sigma_{2}$ defined on the space of the
expansion coefficients:
\begin{equation}
\sigma_{2}(r_{1},r_{2},c_{2})=|c_{2}|+|r_{2}|+c_{1}|r_{1}|+r_{1}^{2},
\label{sigma}
\end{equation}
which measures the degree of cancellation in the expression for
$\rho_{2}$. An unnatural renormalization scheme, which
artificially introduces large expansion coefficients, would be
immediately distinguished by a value of $\sigma_{2}$ which would
be large compared to $|\rho_{2}|$. The function $\sigma_{2}$
defines classes of equivalence of the perturbative approximants.
If one has any preference for using a perturbative expression
obtained in some optimal scheme, one should also take into
account predictions obtained in the schemes which imply the same,
or smaller, cancellations in the expression for $\rho_{2}$, i.e.
which have the same, or smaller, value of $\sigma_{2}$. In
particular, for the PMS scheme we have
$\sigma_{2}\approx2|\rho_{2}|$ \cite{racz95a}. Therefore it
appears that the set of schemes which generate approximants
satisfying $\sigma_{2}\leq2|\rho_{2}|$ is a minimal set that has
to be taken into account in the analysis of stability of the
predictions with respect to change of the RS. More generally, it
is useful to use a condition on the allowed schemes in the form:
\begin{equation}
\sigma_{2}(r_{1},r_{2},c_{2}) \leq l\,|\rho_{2}|,
\label{constraint}
\end{equation}
where $l\geq 1$ is some constant, which determines how strong
cancellations in the expression for $\rho_{2}$ we want to allow.

In this note we analyze the RS dependence of the NNLO predictions
for $\delta_{e^{+}e^{-}}$, using systematically the condition
(\ref{constraint}). As in the previous papers
\cite{racz92,racz93,racz95a}, we use the $r_{1}$ and $c_{2}$
coefficients as the two independent parameters characterizing the
freedom of choice of the approximants in the NNLO. To obtain the
numerical value of the running coupling constant we use the
implicit equation, which results from integrating the
renormalization group equation (\ref{rge}) with appropriate
boundary condition \cite{bard78}:
\begin{equation}
\frac{b}{2}\ln\left(\frac{s}
{\Lambda^{2}_{\overline{MS}}}\right)=
r^{\overline{MS}}_{1}-r_{1}+\Phi(a,c_{2}),
\label{intrge}
\end{equation}
where
\begin{equation}
\Phi(a,c_{2})=c_{1}\ln\left(\frac{b}{2c_{1}}\right)+
\frac{1}{a}+c_{1}\ln(c_{1}a)+O(a).
\end{equation}
The explicit form of $\Phi(a,c_{2})$ is given for example in
\cite{maxw83}. The appearance of $\Lambda_{\overline{MS}}$ and
$r^{\overline{MS}}_{1}$ in the expression (\ref{intrge}) is a
result of taking into account the so called Celmaster-Gonsalves
relation \cite{celm79b} between the lambda parameters in different
schemes. This relation is valid to all orders of perturbation
expansion.

The region of the scheme parameters satisfying
Eq.~(\ref{constraint}) has simple analytic description. In the
case $\rho_{2}<0$ and $|\rho_{2}|>2c_{1}^{2}(l+1)/(l-1)^{2}$ let
us define:
\begin{eqnarray}
r_{1}^{min}&=& - \sqrt{|\rho_{2}|(l+1)/2},\\
r_{1}^{max}&=&[-c_{1}+\sqrt{c_{1}^{2}+2(l+1)|\rho_{2}}|\;]/2,\\
c_{2}^{min}&=&-|\rho_{2}|(l+1)/2,\\
c_{2}^{max}&=&|\rho_{2}|(l-1)/2,\\
c_{2}^{int}&=&c_{1}r_{1}^{min}+c_{2}^{max}.
\end{eqnarray}
For $c_{2}>0$ the region of allowed parameters is bounded from
 above by the line joining the points $(r_{1}^{min},0)$,
 $(r_{1}^{min},c_{2}^{int})$, $(0,c_{2}^{max})$,
 $(r_{1}^{max},c_{2}^{max})$, $(r_{1}^{max},0)$. For $c_{2}<0$
 the region of allowed parameters is bounded from below by the
 lines:
\begin{eqnarray}
c_{2}(r_{1})&=&r_{1}^{2}+c_{2}^{min}\,\,\,\,\,\,\,\,\,\,\,
\mbox{for}\,\,\,\,r_{1}^{min}\leq r_{1}\leq 0,\\
c_{2}(r_{1})&=&r_{1}^{2}+c_{1}r_{1}+c_{2}^{min}\,\,\,\,
\mbox{for}\,\,\,\,0\leq r_{1}\leq r_{1}^{max}.
\end{eqnarray}

In the case $\rho_{2}<0$ and
$|\rho_{2}|<2c_{1}^{2}(l+1)/(l-1)^{2}$ we should use instead:
\begin{eqnarray}
r_{1}^{min}&=&-|\rho_{2}|(l-1)/(2c_{1}),\\
c_{2}^{int}&=&(r_{1}^{min})^{2}+c_{2}^{min}.\\
\end{eqnarray}
For $c_{2}>0$ the region of allowed parameters is then bounded
 from above by the line joining the points $(r_{1}^{min},0)$,
 $(0,c_{2}^{max})$, $(r_{1}^{max},c_{2}^{max})$,
 $(r_{1}^{max},0)$. For $c_{2}<0$ the region of allowed
 parameters is bounded from below by the line joining the points
 $(r_{1}^{min},0)$ and $(r_{1}^{min},c_{2}^{int})$, and the
 curves defined in the previous case.

For $\rho_{2}>c_{1}^{2}/4$ the region of the scheme parameters
satisfying the Eq.~(\ref{constraint}) has been described in
\cite{racz95a}.

\section{Estimate of the RS ambiguities in the conventional
expansion for $\delta_{e^{+}e^{-}}$}

Let us first consider the case of five active quark flavors,
which is most important for experimental determination of
$\Lambda_{\overline{MS}}$. The same corrections gives also a
dominant QCD contribution to the hadronic width of the $Z^{0}$
boson. For $n_{f}=5$ we have $\rho_{2}^{R}=-15.09262$. In Fig.~1
we show the contour plot of $\delta^{(2)}_{e^{+}e^{-}}$ as a
function of the parameters $r_{1}$ and $c_{2}$, for
$\sqrt{s}/\Lambda^{(5)}_{\overline{MS}}=75$. We have indicated
the region of parameters satisfyings the condition
(\ref{constraint}) with $l=2$. For comparison, we also indicate
the region corresponding to $l=3$. The PMS prediction is
represented in Fig.~1 by a saddle point at $r_{1}=-0.408$ and
$c_{2}=-23.154$. We see that the PMS parameters are close to the
approximate solution of the PMS equations \cite{penn82}:
\begin{equation} 
r_{1}^{PMS}=0(a^{PMS}),\;\;\;\;\;\;\;\;
c_{2}^{PMS}=\frac {3}{2} \rho_{2} + 0(a^{PMS}),
\label{aPMS}
\end{equation} 
and the PMS point lies indeed on the boundary of the $l=2$
region, as expected. Comparing the values of
$\delta^{(2)}_{e^{+}e^{-}}$ obtained for the scheme parameters in
the $l=2$ region we find for
$\sqrt{s}/\Lambda^{(5)}_{\overline{MS}}=75$, that the minimal
value is attained for $r_{1}=-4.76$, $c_{2}=1.55$ and the maximal
value is attained for $r_{1}=3.52$, $c_{2}=7.55$. For the $l=3$
region we obtain the minimal value for $r_{1}=-5.49$,
$c_{2}=8.17$, and the maximal value for $r_{1}=3.98$,
$c_{2}=15.09$. In both cases the maximal and minimal values are
attained at the boundary of the allowed region. Let us note, that
the commonly used $\overline{\mbox{MS}}$ scheme lies within the
$l=2$ region.

Performing similar contour plots in the range
$40<\sqrt{s}/\Lambda^{(5)}_{\overline{MS}}<200$ we find, that the
scheme parameters, for which $\delta^{(2)}_{e^{+}e^{-}}$ reaches
extremal values in the $l=2,3$ allowed regions, are practically
independent of the $\sqrt{s}/\Lambda^{(5)}_{\overline{MS}}$.

In Fig.~2 we show how the maximal and minimal values of
$\delta^{(2)}_{e^{+}e^{-}}$ in the $l=2,3$ allowed regions depend
on $\sqrt{s}/\Lambda^{(5)}_{\overline{MS}}$. We also show the PMS
prediction and the experimental constraint
$\delta^{exp}_{e^{+}e^{-}}(\sqrt{s}=31.6\,\mbox{GeV})=0.0527\pm0.0050$
\cite{mars89}. We find that with increasing
$\sqrt{s}/\Lambda^{(5)}_{\overline{MS}}$ the scheme dependence is
decreasing, as expected, although it remains substantial even for
high energies. Let us take for example
$\sqrt{s}/\Lambda^{(5)}_{\overline{MS}}=162$, corresponding to
$\Lambda^{(5)}_{\overline{MS}}=0.195\,\mbox{GeV}$ --- which is
the value preferred by the Particle Data Group \cite{pdgr94} ---
and $\sqrt{s}=31.6$. In this case the scheme variation of
$\delta^{(2)}_{e^{+}e^{-}}$ over the $l=2$ region is $5\%$ of the
PMS prediction, and for the $l=3$ region $8\%$, compared with
$9\%$ and $16\%$, respectively, for
$\sqrt{s}/\Lambda^{(5)}_{\overline{MS}}=75$. However, when we
decrease $\sqrt{s}/\Lambda^{(5)}_{\overline{MS}}$ below 75 the
scheme dependence increases rapidly, and it becomes very large
already for $\sqrt{s}/\Lambda^{(5)}_{\overline{MS}}=30$. The
scheme dependence appears to be quite large in the range of
values of $\delta^{(2)}_{e^{+}e^{-}}$ relevant for fitting the
experimental data. For example, the line representing the minimal
values on the $l=2$ region does not reach the central
experimental value, which translates into a very large
theoretical uncertainty in the fitted value of
$\Lambda_{\overline{MS}}$.

For $n_{f}=4$ we have $\rho_{2}^{R}=-13.30991$. In Fig.~3 we show
 the contour plot of $\delta^{(2)}_{e^{+}e^{-}}$ as a function of
 the parameters $r_{1}$ and $c_{2}$, for
 $\sqrt{s}/\Lambda^{(4)}_{\overline{MS}}=30$. Similarly as in the
 $n_{f}=5$ case we find that the PMS prediction is well
 represented by the approximate solution (\ref{aPMS}). The
 variation over the $l=2$ region is approximately $11\%$ of the
 PMS prediction. In Fig.~4 we show the variation in the
 predictions for $\delta^{(2)}_{e^{+}e^{-}}$ when the scheme
 parameters are changed over the $l=2$ region, as a function of
 $\sqrt{s}/\Lambda^{(4)}_{\overline{MS}}$. It is evident that for
 $\sqrt{s}/\Lambda^{(4)}_{\overline{MS}}$ smaller that 20, which
 is the range relevant for fitting the experimental data, this
 variation becomes very large. (Note that analysis of
 experimental data from several experiments gives \cite{mars89}
 $\delta^{exp}_{e^{+}e^{-}}(\sqrt{s}=9\,\mbox{GeV})=0.073\pm0.024$.)

Finally, for $n_{f}=3$ we have $\rho_{2}^{R}=-11.41713$. In
 Fig.~5 we show the contour plot of $\delta^{(2)}_{e^{+}e^{-}}$
 as a function of the parameters $r_{1}$ and $c_{2}$, for
 $\sqrt{s}/\Lambda^{(3)}_{\overline{MS}}=9$. The variation of the
 predictions over the $l=2$ region is approximately $28\%$ of the
 PMS value. In Fig.~6 we show the variation in the predictions
 for $\delta^{(2)}_{e^{+}e^{-}}$ when the scheme parameters are
 changed over the $l=2$ region, as a function of
 $\sqrt{s}/\Lambda^{(3)}_{\overline{MS}}$. We observe that the
 variation in the predictions starts to increase rapidly for
 $\sqrt{s}/\Lambda^{(3)}_{\overline{MS}}$ smaller than 9.

Let us summarize our analysis of the predictions for
$\delta_{e^{+}e^{-}}$ obtained from the conventional expansion.
We found that changing the renormalization scheme within a class
of schemes which, according to our condition (\ref{constraint}),
appear to be as good as the PMS scheme, we obtain rather large
variation in the predictions. In some cases we may even speak
about instability of the predictions with respect to change of
the renormalization scheme. This is in contrast with the
statement in \cite{chyl91}, that the conventional expansion for
$\delta_{e^{+}e^{-}}$ is highly reliable. The conclusion found in
\cite{chyl91} is based on the observation, that for
$\delta_{e^{+}e^{-}}$ the $\overline{\mbox{MS}}$ prediction is
very close to the PMS prediction. The fact that the
$\overline{\mbox{MS}}$ prediction is very close to the PMS
prediction is of course true --- for example in the scale of
Fig.~2 the $\overline{\mbox{MS}}$ and PMS curves would be
difficult to distinguish. Similar situation occurs for other
values of $n_{f}$. It is clear however, that there is no
theoretical or phenomenological motivation to use the
$\overline{\mbox{MS}}$-PMS difference as a measure of reliability
of the perturbation expansion for any physical quantity. The fact
that the $\overline{\mbox{MS}}$ prediction for
$\delta_{e^{+}e^{-}}$ is close to the PMS prediction is simply a
coincidence, without deeper significance for such problems as
reliability of the predictions and good or bad convergence of the
perturbation expansion.

It is interesting to note that for very low energies the PMS
 predictions display the infrared fixed point type of behavior
 \cite{matt92}. However, this type of behavior, which in fact does
 not manifest itself in the $n_{f}=3$ predictions until
 $\sqrt{s}/\Lambda^{(3)}_{\overline{MS}}\approx2.5$, is
 accompanied by a rapidly increasing RS dependence. It seems
 therefore unreasonable to put too much faith in the PMS
 prediction when even a very small change of the scheme parameters
 dramatically modifies the result. These remarks apply as well to
 the case $n_{f}=2$.

\section{Analysis of the $\pi^{2}$ terms in
$\delta_{e^{+}e^{-}}$}

The strong RS dependence described above is somewhat surprising.
It may seem understandable that the perturbation expansion is not
reliable in the energy range appropriate for example for the
$n_{f}=3$ regime. However, one would expect that
$\sqrt{s}/\Lambda^{(5)}_{\overline{MS}}$ of order 75 is large
enough for the perturbation series to be very well behaved. The origin
of the strong scheme dependence may be traced back to the fact
that the NNLO correction is relatively large, which is reflected
by large value of the RS invariant $\rho_{2}^{R}$. However, a
major contribution to the NNLO correction comes from the term
which appears in the process of analytic continuation of
perturbative expression from spacelike to timelike momenta. To
see clearly the significance of such contributions, and to show
how one may treat them in an improved way, it is convenient to
use the so called Adler function \cite{adle74}:
\begin{equation}
D(q^{2})=-12\pi^{2}\,q^{2}\frac{d\,}{d q^{2}}\Pi(q^{2}).
\label{adler}
\end{equation}
where $\Pi(q^{2})$ is the transverse part of quark
electromagnetic current correlator $\Pi^{\mu\,\nu}(q)$:
\begin{eqnarray}
\Pi^{\mu\nu}(q)&=&(-g^{\mu\nu}q^{2}+q^{\mu}q^{\nu})\,\Pi(q^{2}),\\
\Pi^{\mu\nu}(q)&=&i\int\,d^{4}x\,e^{iqx}\,<0|T(J^{\mu}(x)
J^{\nu}(0)^{\dagger})|0>.
\end{eqnarray}
Neglecting the quark mass effects and electroweak corrections we
may write:
\begin{equation}
D(q^{2})= 3\,\sum_{f}\,Q^{2}_{f}\,[ 1 + \delta_{D}(-q^{2})],
\end{equation}
where $\delta_{D}(-q^{2})$ denotes the QCD correction. The Adler
function is RS invariant in the formal sense, i.e. it may be
considered to be a physical quantity, despite the fact that it
cannot be directly measured in the experiment. In particular,
$\delta_{D}(-q^{2})$ is renormalization group invariant, in contrast
to $\Pi(q^{2})$, which does not even satisfy a homogenous
renormalization group equation \cite{nari82}.  The Adler function
is directly 
calculable in the perturbation expansion for spacelike momenta.
To express the $R_{e^{+}e^{-}}$ ratio by the Adler function one
inverts the relation (\ref{adler}):
\begin{equation}
\Pi(q^{2})-\Pi(q_{0}^{2})=-\,\frac{1}{12\pi^{2}}
\int_{q_{0}^{2}}^{q^{2}}\,d\sigma\,\frac{D(\sigma)}{\sigma},
\end{equation}
where $q_{0}^{2}$ is some reference spacelike momentum, and one
utilizes the relation:
\begin{equation}
R_{e^{+}e^{-}}(s)=12\,\pi\,\mbox{Im}\Pi(s+i\epsilon)=
\frac{6\pi}{i}\left[\Pi(s+i\epsilon)-\Pi(s-i\epsilon)\right].
\end{equation}
In this way one obtains $R_{e^{+}e^{-}}(s)$ as a contour integral
in the complex momentum plane, with the Adler function under the
integral \cite{radu82,radu86}:
\begin{equation}
R_{e^{+}e^{-}}(s)=-\,\frac{1}{2\pi i}\int_{C}\,
d\sigma\,\frac{D(\sigma)}{\sigma},
\end{equation}
where the contour C runs clockwise from $\sigma=s-i\epsilon$ to
$\sigma=0$ below the real positive axis, around $\sigma=0$, and
to $\sigma=s+i\epsilon$ above the real positive axis. The
integration contour may be of course arbitrarily deformed in the
domain of analyticity of the Adler function. A convenient choice
is $q^{2}=-s\exp (-i\theta)$ with $\theta\in[-\pi,\pi]$. For this
choice of the contour we obtain the following simple relation
between $\delta_{e^{+}e^{-}}(s)$ and $\delta_{D}(-q^{2})$:
\begin{equation}
\delta_{e^{+}e^{-}}(s)=\frac{1}{2\pi}\,\int_{-\pi}^{\pi}\,d\theta
\,\left[\delta_{D}(-\sigma)|_{\sigma=-s\exp (-i\theta)}\right],
\label{dcont}
\end{equation}
The conventional expression for $\delta_{e^{+}e^{-}}(s)$ may be
recovered from this formula by inserting under the contour
integral an expansion of $\delta_{D}(-q^{2})$ in terms of $a(s)$:
\begin{eqnarray}
\lefteqn{
\delta^{(2)}_{D}(-q^{2})=a(s)
\left\{
1+\left[\hat{r}_{1}-(b/2)\ln(-q^{2}/s)\right]\,a(s)+
\right.
}\nonumber\\
& & +\left.
\left[\hat{r}_{2}-(c_{1}+2\,\hat{r}_{1})(b/2)\ln(-q^{2}/s)+
(b/2)^{2}\,(\ln(-q^{2}/s))^{2}\right]\,a^{2}(s)\right\},
\end{eqnarray}
where $\hat{r}_{i}$ denote the coefficients for expansion of
$\delta_{D}(-q^{2})$ in terms of $a(-q^{2})$. Evaluating the
trivial contour integrals involving powers of $\ln(-\sigma/s)$,
we obtain the expression (\ref{delta}) with:
\begin{equation}
r_{1}=\hat{r}_{1},\,\,\,\,\,
r_{2}=\hat{r}_{2}-\frac{1}{3}\,\left(\frac{b\pi}{2}\right)^{2}.
\end{equation}
This implies $\rho_{2}^{R}=\rho_{2}^{D}-(b\pi/2)^{2}/3$. In
Table~2 we list the values of $\rho_{2}^{D}$ for various values
of $n_{f}$.

\begin{table}
\begin{center}
\begin{tabular}{||c|c||}
\hline
$n_{f}$ 
    &$\rho_{2}^{D}$\\
\hline
2&  9.28877\\
\hline
3&  5.23783\\
\hline
4&  0.96903\\
\hline
5& -3.00693\\
\hline
6& -7.36281\\
\hline
\end{tabular}   
\end{center}
\caption{ Numerical values of the RS invariant $\rho_{2}^{D}$
 characterisitic for the QCD correction to the Adler function.}
\end{table}
Numerically the contribution of the $\pi^{2}$ term is very large
--- for example for $n_{f}=5$ we have
$\rho_{2}^{R}-\rho^{D}_{2}=-12.08570$.

Contributions proportional to $\pi^{2}$ appear also in higher
orders. We have \cite{bjor89}:
\begin{eqnarray}
r_{3}&=&\hat{r}_{3}-\left(\hat{r}_1+\frac{5}{6}\,c_{1}\right)\,
\left(\frac{b\pi}{2}\right)^{2},
\label{pi2corr4}\\
r_{4}&=&\hat{r}_{4}-\,\left(2\hat{r}_{2}+\frac{7}{3}c_{1}\hat{r}_{1}+
\frac{1}{2}c^{2}_{1}+c_{2}\right)\,\left(\frac{b\pi}{2}\right)^{2}+
\frac{1}{5}\left(\frac{b\pi}{2}\right)^{4},
\label{pi2corr5}
\end{eqnarray}
The result for $r_{5}$ may be found in \cite{kata95}:
\begin{eqnarray}
r_{5}&=&\hat{r}_{5}-\frac{1}{3}\,\left(10\hat{r}_{3}+
\frac{27}{2}c_{1}\hat{r}_{2}+4c^{2}_{1}\hat{r}_{1}+
\frac{7}{2}c_{1}c_{2}+8c_{2}\hat{r}_{1}+\frac{7}{2}c_{3}\right)\,
\left(\frac{b\pi}{2}\right)^{2}+
\nonumber\\
& & +\frac{1}{5}\left(5\hat{r}_{1}+\frac{77}{12}c_{1}\right)\,
\left(\frac{b\pi}{2}\right)^{4}.
\label{pi2corr6}
\end{eqnarray}
(The difference between $r_{i}$ and $\hat{r}_{i}$ in higher
orders was studied in \cite{bjor89,brow92,pivo92a}.) Note that
the $\pi^{2}$ corrections to $\hat{r}_{3}$ and $\hat{r}_{4}$ are
fully determined by the NNLO expression for $\delta_{D}(-q^{2})$.
Taking into account that we have the following expressions for
the higher order RS invariant combinations of the expansion
coefficients \cite{dhar83}:
\begin{eqnarray}  
\rho_{3}&=&c_{3}+2r_{3}-4r_{2}r_{1}-2r_{1}\rho_{2}-
r_{1}^{2}c_{1}+2r^{3}_{1},\\
\rho_{4}&=&c_{4}+3r_{4}-6r_{3}r_{1}-4r^{2}_{2}-3r_{1}\rho_{3}-
4r^{4}_{1}-\nonumber\\
& & -(r_{2}+2r^{2}_{1})\rho_{2}+11r_{2}r^{2}_{1}
+c_{1}(r_{3}-3r_{2}r_{1}+r^{3}_{1}). 
\end{eqnarray}
we obtain:
\begin{eqnarray}
\rho_{3}^{R}&=&\rho^{D}_{3}-
\frac{5}{3}c_{1}\,\left(\frac{b\pi}{2}\right)^{2},\\
\rho_{4}^{R}&=&\rho^{D}_{4}-\frac{1}{3}(8\rho^{D}_{2}+7c^{2}_{1})
\,\left(\frac{b\pi}{2}\right)^{2}+
\frac{2}{45}\left(\frac{b\pi}{2}\right)^{4}.
\end{eqnarray}
The $\pi^{2}$ terms are quite
sizeable numerically. For example for $n_{f}=5$ we have:
\begin{equation}
\rho_{3}^{R}-\rho^{D}_{3}=-76.1924,\,\,\,
\rho_{4}^{R}-\rho^{D}_{4}=211.025.
\end{equation}
It is evident that the terms arising from the analytic
continuation would make a significant contribution to the RS
invariants in any order of the perturbation expansion.

Returning to the evaluation of $\delta_{e^{+}e^{-}}(s)$, we note
that the procedure used to obtain the conventional result treats
the $q^{2}$ dependence of $\delta_{D}$ in the complex energy
plane in a rather crude way. A straightforward way to improve
this evaluation is to use under the contour integral the
renormalization group improved expression for
$\delta_{D}(-\sigma)$, analytically continued from the real
negative $\sigma$ to the whole complex energy plane cut along the
real positive axis.  In other words, one should take into account
the renormalization group evolution of $a(-\sigma)$ in the complex
energy plane, avoiding the expansion of $a(-\sigma)$ in terms of
$a(s)$. In this way one makes maximal use of the renormalization
group invariance property of the Adler function. Of course the
integral may be now done only numerically, and the resulting
expression for $\delta_{e^{+}e^{-}}(s)$ is no longer a polynomial
in $a(s)$, despite the fact that only the NNLO expression for the
Adler function is used. It is easy to convince onself that the
procedure outlined above is equivalent to the resummation --- to all
orders --- of the $\pi^{2}$ terms 
that contain powers of $b$, $c_{1}$ and/or $c_{2}$.
(The summation of the leading terms
proportional to $(b\pi/2)^{2}$ was discussed in \cite{pivo92a}.)

The improved approach 
based on the contour integral has been implemented with success
in the case of the QCD corrections to the tau lepton decay
\cite{pivo92b,ledi92,racz94}, where a similar problem of strong
renormalization scheme dependence appears. It was found that
using the contour integral representation and evaluating the
contour integral numerically one obtains considerable improvement
in the stability of predictions with respect to change of RS
\cite{ledi92,racz94}. It 
is therefore of great interest to see whether one may improve in
this way the predictions for $\delta_{e^{+}e^{-}}$.

\section{Improved evaluation of $\delta_{e^{+}e^{-}}$}

In this section we perform an analysis similar to that in the
Section~3, using now the improved predictions for
$\delta_{e^{+}e^{-}}$, obtained by evaluating numerically the
contour integral in Eq.~(\ref{dcont}). Similarly as in the case
of the conventional perturbation expansion, we begin with the
$n_{f}=5$ case. To show, how the improved evaluation of
$\delta^{(2)}_{e^{+}e^{-}}$ affects its RS dependence, we compare
the plots of $\delta^{(2)}_{e^{+}e^{-}}$ as a function of
$r_{1}$, for several values of $c_{2}$, with
$\sqrt{s}/\Lambda^{(5)}_{\overline{MS}}=75$, obtained with the
conventional NNLO expression (Fig.~7) and with the numerical
evaluation of the contour integral (Fig.~8). We see that the
predictions obtained by the numerical evaluation of the contour
integral have much smaller RS dependence. In Fig.~8 we have also
indicated the predictions obtained with the conventional
expansion supplemented by the $O(a^{4})$ and $O(a^{5})$ terms
given by Eq.~(\ref{pi2corr4}) and Eq.~(\ref{pi2corr5}). We see
that this type of simple 
improvement of the conventional expansion reproduces quite well
the results obtained from exact contour integral, except for
large negative $r_{1}$. (Inclusion of only the $O(a^{4})$ term
does not give good approximation. Inclusion of the $O(a^{6})$
correction given by Eq.~(\ref{pi2corr6}), which is of course only
partially known at present, slightly improves the approximation
for positive $r_{1}$.)

 In Fig.~9 we show the contour plot of
$\delta^{(2)}_{e^{+}e^{-}}$ obtained from the expression
(\ref{dcont}) for $\sqrt{s}/\Lambda^{(5)}_{\overline{MS}}=75$. In
Fig.~9 we also show the relevant regions of the scheme parameters
satisfying the condition (\ref{constraint}) with $l=2,3$. These
regions are calculated assuming $\rho_{2}=\rho_{2}^{D}$, because
the basic object in the improved approach is $\delta^{(2)}_{D}$.
For $n_{f}=5$ we have $\rho_{2}^{D}=-3.00693$, which is much
smaller in absolute value than $\rho^{R}$. Consequently, the
region of the allowed scheme parameters is much smaller than in
the analysis of the conventional NNLO approximant. The improved
predictions for $\delta_{e^{+}e^{-}}$  have a saddle point
type of behavior as a function of $r_{1}$ and $c_{2}$, where the
saddle point represents the PMS prediction. However, the location
of the saddle point is completely different than in the case of
conventional expansion. (The location of the saddle point for the
improved expression is no longer a solution of the set of the PMS
equations given in \cite{stev81}, because the improved
approximant (\ref{dcont}) is not a polynomial in the running
coupling constant.) It is interesting that the PMS point for the
improved expression lies very close to the point $r_{1}=0$ and
$c_{2}=1.5\rho^{D}_{2}=-4.51$, which corresponds to the
approximate value of the PMS parameters if $\delta^{(2)}_{D}$ is
optimized for spacelike momenta. Let us note that the
$\overline{\mbox{MS}}$ scheme lies outside the $l=2$ region in
this case. However, the $\overline{\mbox{MS}}$ prediction in the
improved approach is very close to the improved PMS prediction:
we have 0.05279 and 0.05275 respectively. 

The variation of the
predictions over the $l=2$ region is $0.3\%$ of the PMS
prediction, and variation over the $l=3$ region is $0.5\%$ of the
PMS prediction. Even if we take variation over the region
corresponding to $l=10$ we obtain only $2.5\%$ change in the
predictions. We see that the improved prediction for
$\delta_{e^{+}e^{-}}$ shows wonderful stability with respect to
change of the RS. From Fig.~7 and Fig.~8 it is also clear, that
the difference between NNLO and NLO PMS predictions is much
smaller in the case of the improved prediction --- $0.9\%$ of the
NNLO result for $\sqrt{s}/\Lambda^{(5)}_{\overline{MS}}=75$ ---
than in the case of the conventional expansion --- $4.7\%$ of the
NNLO result. We conclude therefore that the theoretical
ambiguities involved in the evaluation of
$\delta^{(2)}_{e^{+}e^{-}}$ are in fact very small, provided that
the analytic continuation effects are treated with appropriate
care. For completeness, we give in Table~3 the  NNLO and
NLO PMS predictions in the improved approach for several values of
$\sqrt{s}/\Lambda^{(5)}_{\overline{MS}}$.

\begin{table}
\begin{center}
\begin{tabular}{||c|c|c||}
\hline
$\sqrt{s}/\Lambda^{(5)}_{\overline{MS}}$
              &$\delta_{e^{+}e^{-}}^{opt,NNLO}$
                       &$\delta_{e^{+}e^{-}}^{opt,NLO}$\\
\hline
  25   & 0.06799 & 0.06888\\
\hline
  50   & 0.05753 & 0.05811\\
\hline
  75   & 0.05275 & 0.05320\\
\hline
100   & 0.04981 & 0.05019\\
\hline
200   & 0.04389 & 0.04415\\
\hline
500   & 0.03791 & 0.03809\\
\hline
\end{tabular}   
\end{center}
\caption{ Numerical values of the optimized predictions for
$\delta_{e^{+}e^{-}}$, obtained from the contour integral
expression (\protect\ref{dcont}) for $n_{f}=5$. The PMS
parameters are well approximated by $r_{1}=0$,
$c_{2}=1.5\rho_{2}^{D}$ (NNLO) and $r_{1}=-0.59$ (NLO).} 
\end{table}

In the case of $n_{f}=5$ predictions it is interesting how the
improved evaluation affects the fit to experimental data. Using
the experimental constraint
$\delta^{exp}_{e^{+}e^{-}}(\sqrt{s}=31.6\,\mbox{GeV})=0.0527\pm0.0050$
\cite{mars89} and the improved PMS prediction we find
$\Lambda^{(5)}_{\overline{MS}}=0.419\pm0.194\,\mbox{GeV}$, which
is equivalent in the three loop approximation to
$\alpha_{s}^{\overline{MS}}(M_{Z}^{2})=0.1319\pm0.0100$. For
comparison, using the conventional expansion in the
$\overline{\mbox{MS}}$ scheme we obtain the central value of
$\Lambda^{(5)}_{\overline{MS}}=0.399\,\mbox{GeV}$
($\alpha_{s}^{\overline{MS}}(M_{Z}^{2})=0.1308$), while with the
PMS prescription in the conventional expansion we get
$\Lambda^{(5)}_{\overline{MS}}=0.410\,\mbox{GeV}$
($\alpha_{s}^{\overline{MS}}(M_{Z}^{2})=0.1314$). We see
therefore that improvement in the evaluation of
$\delta^{(2)}_{e^{+}e^{-}}$ has  small effect on the fitted
values of the $\Lambda^{(5)}_{\overline{MS}}$ parameter.

For $n_{f}=4$ we have $\rho^{D}_{2}=0.96903$, i.e. the effect of
$\pi^{2}$ corrections is even larger than in the $n_{f}=5$ case.
The $n_{f}=4$ case is in all respects similar to the $n_{f}=5$
case, except for the fact that the reduction in RS dependence
seems to be even stronger. In Fig.~10 and Fig.~11 we compare the
plots of $\delta^{(2)}_{e^{+}e^{-}}$ as a function of $r_{1}$, for
several values of $c_{2}$, with
$\sqrt{s}/\Lambda^{(4)}_{\overline{MS}}=30$, obtained with the
conventional NNLO expression (Fig.~10) and with the numerical
evaluation of the contour integral (Fig.~11). In Fig.~11 we also
show the predictions obtained with the conventional expansion
supplemented by the $O(a^{4})$ and $O(a^{5})$ terms given by
Eq.~(\ref{pi2corr4}) and Eq.~(\ref{pi2corr5}). (Inclusion of the
$O(a^{6})$ correction 
(\ref{pi2corr6}) does not improve the approximation.) In Fig.~12
we show the contour plot of the improved prediction for
$\delta_{e^{+}e^{-}}$ obtained for
$\sqrt{s}/\Lambda^{(4)}_{\overline{MS}}=30$. It is interesting
that variation of the predictions over the $l=2$ region is
extremely small, of the order of $0.03\%$ (!) of the PMS
prediction. The improved prediction for
$\sqrt{s}/\Lambda^{(4)}_{\overline{MS}}=30$   in the
$\overline{\mbox{MS}}$ 
scheme is 0.05902, quite close to the improved PMS result
0.05907. The differences with the results obtained in the
conventional approach again are not very big --- using the
conventional expansion we have 0.06025 in the
$\overline{\mbox{MS}}$ scheme and 0.05975 in the NNLO PMS. In
Table~4 we give numerical values of the improved predictions in
the PMS scheme, for several values of
$\sqrt{s}/\Lambda^{(4)}_{\overline{MS}}$. We find that in the
improved approach the NNLO PMS predictions are very close to NLO
PMS predictions. We see therefore that also for $n_{f}=4$ the
theoretical uncertainties in the improved predictions for
$\delta_{e^{+}e^{-}}$ are very small.

\begin{table}
\begin{center}
\begin{tabular}{||c|c|c||}
\hline
$\sqrt{s}/\Lambda^{(4)}_{\overline{MS}}$
              &$\delta_{e^{+}e^{-}}^{opt,NNLO}$
                       &$\delta_{e^{+}e^{-}}^{opt,NLO}$\\
\hline
 10   & 0.08108 & 0.08093\\
\hline
 20   & 0.06574 & 0.06565\\
\hline
 30   & 0.05907 & 0.05900\\
\hline
 40   & 0.05508 & 0.05503\\
\hline
 50   & 0.05233 & 0.05228\\
\hline
\end{tabular}   
\end{center}
\caption{Same as in Table 3, but for $n_{f}=4$. The PMS parameter
in NLO is approximately $r_{1}=-0.71$.}

\end{table}

Finally let us consider the case of $n_{f}=3$. We have then
$\rho^{D}_{2}=5.23783$. In Fig.~13 and Fig.~14 we compare the
plots of $\delta^{(2)}_{e^{+}e^{-}}$ as a function of $r_{1}$, for
several values of $c_{2}$, with
$\sqrt{s}/\Lambda^{(3)}_{\overline{MS}}=9$, obtained with the
conventional NNLO expression (Fig.~13) and with the numerical
evaluation of the contour integral (Fig.~14). Again, we find
dramatic reduction in the RS dependence, despite rather low
energy. It is interesting that 
in the $n_{f}=3$ case the addition of $\pi^{2}$ corrections given
by Eq.~(\ref{pi2corr4}) and  Eq.~(\ref{pi2corr5}) does not result
in the improvement of the 
conventional predictions. In Fig.~15 we show the contour plot of
$\delta^{(2)}_{e^{+}e^{-}}$ obtained from the improved expression for
$\sqrt{s}/\Lambda^{(3)}_{\overline{MS}}=9$. Similarly as for
other numbers of flavors we obtain in the improved approach a
very small variation in the predictions when parameters are
changed over the $l=2$ region of parameters appropriate for
$\delta^{(2)}_{D}$ --- the variation is of the order of $0.8\%$ of the
PMS prediction 0.07756. (We have verified that this situation
persists down to $\sqrt{s}/\Lambda^{(3)}_{\overline{MS}}=4$.) The
improved prediction in the $\overline{\mbox{MS}}$ scheme is
0.07719. For comparison, in the conventional approach we obtain
0.08097 in the NNLO PMS and 0.08244 in the NNLO
$\overline{\mbox{MS}}$ scheme. In Table~5 we give numerical
values of the improved predictions in the PMS scheme, for several
values of $\sqrt{s}/\Lambda^{(3)}_{\overline{MS}}$. With this
results we conclude, that the $n_{f}=3$ NNLO expression for
$\delta_{e^{+}e^{-}}$, obtained by evaluating the contour
integral (\ref{dcont}) numerically, has very small theoretical
uncertainty, even for rather small values of
$\sqrt{s}/\Lambda^{(3)}_{\overline{MS}}$. This situation is
similar to that found for the QCD corrections to the tau decay
\cite{ledi92,racz94}.

The behavior of $\delta^{(2)}_{e^{+}e^{-}}$ at very low energies and
the problem of existence of the fixed point in the improved
approach would be discussed in a separate note \cite{racz96}.

\begin{table}
\begin{center}
\begin{tabular}{||c|c|c||}
\hline
$\sqrt{s}/\Lambda^{(3)}_{\overline{MS}}$
               &$\delta_{e^{+}e^{-}}^{opt,NNLO}$
                       &$\delta_{e^{+}e^{-}}^{opt,NLO}$\\
\hline
  5 & 0.09624 & 0.09421\\
\hline
  7 & 0.08475 & 0.08312\\
\hline
  9 & 0.07756 & 0.07619\\
\hline
 11 & 0.07255 & 0.07136\\
\hline
 13 & 0.06879 & 0.06774\\
\hline
\end{tabular}   
\end{center}
\caption{ Same as in Table~3, but for $n_{f}=3$. The
PMS parameter in NLO is approximately $r_{1}=-0.81$.}
\end{table}

\section{Summary and conclusions}

Summarizing, we have analyzed the RS dependence of the
conventional NNLO expression for $\delta_{e^{+}e^{-}}$ using a
systematic method described in \cite{racz92,racz93,racz95a}. We found
rather large variation in the predictions. We have also
investigated an improved way of calculating
$\delta^{(2)}_{e^{+}e^{-}}$, which relies on a contour integral
representation for this quantity and a numerical evaluation of
the contour integral. We found that the stability of 
$\delta^{(2)}_{e^{+}e^{-}}$ with respect to change of
the RS is greatly improved when the contour integral approach is
used.  Also,
in the improved approach the difference between optimized NNLO
and NLO predictions was found to be much smaller than in the case
of the conventional expansion. We conclude therefore that the
theoretical uncertainties in the NNLO QCD predictions for
$\delta_{e^{+}e^{-}}$ are very small, even at low energies,
provided that large $\pi^{2}$ terms, arising from analytic
continuation, are treated with due care. We observed that the
optimized predictions for $\delta_{e^{+}e^{-}}$, obtained in the
contour integral approach, lie in general below the predictions
from the optimized conventional expansion. However, for $n_{f}=5$
the change in the fit of $\Lambda^{(5)}_{\overline{MS}}$ to the
experimental result came out to be small.

{\em Note added.} After this paper was completed, a related work was
brought to our attention \cite{soper95}, in which the RS dependence of
the QCD corrections to the total hadronic width of the Z boson is
discussed. In \cite{soper95} it is observed, that by using the contour
integral to resumm the large $\pi^{2}$ contributions one reduces the
scale dependence of the QCD predictions. This result is in agreement
with our observations, since the dominant contribution to
$\Gamma^{had}_{Z}$ comes from expression identical to
$\delta_{e^{+}e^{-}}$.  Let us note that the result reported in
\cite{soper95} was anticipated already in \cite{racz95a}. However, the
approach adopted in \cite{soper95} differs from our approach in
several ways. The authors of \cite{soper95} do not discuss the choice
of the range of scheme parameters used in their analysis. In their
investigation of the conventional expansion for $\Gamma^{had}_{Z}$
they use a smaller range of parameters than the one used above for
$n_{f}=5$. In particular, the PMS parameters are outside the range
considered in \cite{soper95}. In the analysis of improved predictions
for $\Gamma^{had}_{Z}$ the authors of \cite{soper95} limit themselves
to the discussion of the renormalization scale dependence, fixing the
$\beta$-function to the $\overline{MS}$ value.  There is also a
technical difference that authors of \cite{soper95} use approximate
analytic expression for the running coupling constant to integrate
along the contour in the complex energy plane, whereas we use exact
numerical solution of the two or three loop renormalization group
equation.

\newpage
\section*{Figure Captions}
\noindent {\bf Fig.~1} The contour plot of
$\delta^{(2)}_{e^{+}e^{-}}$ as a function of the parameters
$r_{1}$ and $c_{2}$, with $n_{f}=5$, for
$\sqrt{s}/\Lambda^{(5)}_{\overline{MS}}=75$. The region of scheme
parameters satisfying the condition (\ref{constraint}) has been
also indicated for $l=2$ (the smaller region) and for
$l=3$.\\

\noindent {\bf Fig.~2} The maximal and minimal values of
$\delta^{(2)}_{e^{+}e^{-}}$ in the $l=2$ (dash-dotted line) and
$l=3$ (dashed line) allowed regions, with $n_{f}=5$, as a
function of $\sqrt{s}/\Lambda^{(5)}_{\overline{MS}}$. The PMS
prediction is also shown, and the experimental constraint
$\delta^{exp}_{e^{+}e^{-}}(\sqrt{s}=31.6\,\mbox{GeV})=0.0527\pm0.0050$
\cite{mars89} is indicated for comparison.\\

\noindent {\bf Fig.~3} The contour plot of
$\delta^{(2)}_{e^{+}e^{-}}$ as a function of the parameters
$r_{1}$ and $c_{2}$, with $n_{f}=4$, for
$\sqrt{s}/\Lambda^{(4)}_{\overline{MS}}=30$. The region of scheme
parameters satisfying the condition (\ref{constraint}) with $l=2$
has been also indicated. \\

\noindent {\bf Fig.~4} The variation in the predictions for
$\delta^{(2)}_{e^{+}e^{-}}$ when the scheme parameters are
changed over the $l=2$ region, with $n_{f}=4$, as a function of
$\sqrt{s}/\Lambda^{(4)}_{\overline{MS}}$. The upper curve
corresponds to $r_{1}=3.10$ and $c_{2}=6.65$, the lower curve
corresponds to $r_{1}=-4.32$ and $c_{2}=0$. For comparison the
PMS prediction is  shown. \\

\noindent {\bf Fig.~5} Same as in Fig.~3 but for $n_{f}=3$ and
 $\sqrt{s}/\Lambda^{(3)}_{\overline{MS}}=9$.\\

\noindent {\bf Fig.~6} The variation in the predictions for
$\delta^{(2)}_{e^{+}e^{-}}$ when the scheme parameters are
changed over the $l=2$ region, with $n_{f}=3$, as a function of
$\sqrt{s}/\Lambda^{(3)}_{\overline{MS}}$. The upper curve
corresponds to $r_{1}=2.71$ and $c_{2}=5.71$, the lower curve
corresponds to $r_{1}=-3.21$ and $c_{2}=0$. For comparison the
PMS curve is shown.\\ 

\noindent {\bf Fig.~7} $\delta^{(2)}_{e^{+}e^{-}}$ as a function of
$r_{1}$, for several values of $c_{2}$, for $n_{f}=5$ and
$\sqrt{s}/\Lambda^{(5)}_{\overline{MS}}=75$, obtained with the
conventional NNLO expression. For comparison also the NLO
predictions are indicated.\\

\noindent {\bf Fig.~8} $\delta^{(2)}_{e^{+}e^{-}}$ as a function of
$r_{1}$, for several values of $c_{2}$, for $n_{f}=5$ and
$\sqrt{s}/\Lambda^{(5)}_{\overline{MS}}=75$, obtained with the
numerical evaluation of the contour integral. For comparison also
the NLO predictions are indicated, and the predictions obtained
from the conventional expansion supplemented by the $O(a^{4})$
and $O(a^{5})$ corrections given by Eq.~(\ref{pi2corr4}) and
Eq.~(\ref{pi2corr5}).\\ 

\noindent {\bf Fig.~9} Contour plot of $\delta^{(2)}_{e^{+}e^{-}}$
obtained from the improved expression for $n_{f}=5$ and
$\sqrt{s}/\Lambda^{(5)}_{\overline{MS}}=75$. The regions of
scheme parameters satisfying the condition (\ref{constraint})
with $l=2$ (the smaller region) and $l=3$ have been indicated,
assuming $\rho_{2}=\rho_{2}^{D}$. \\

\noindent {\bf Fig.~10} Same as in Fig.~7, but for $n_{f}=4$ and
 $\sqrt{s}/\Lambda^{(4)}_{\overline{MS}}=30$.\\

\noindent {\bf Fig.~11} Same as in Fig.~8, but for $n_{f}=4$ and
 $\sqrt{s}/\Lambda^{(4)}_{\overline{MS}}=30$.\\

\noindent {\bf Fig.~12} Same as in Fig.~9, but for $n_{f}=4$ and
 $\sqrt{s}/\Lambda^{(4)}_{\overline{MS}}=30$. Only the $l=2$
 region has been indicated.\\

\noindent {\bf Fig.~13} Same as in Fig.~7, but for $n_{f}=3$ and
 $\sqrt{s}/\Lambda^{(3)}_{\overline{MS}}=9$.\\

\noindent {\bf Fig.~14} Same as in Fig.~8, but for $n_{f}=3$ and
 $\sqrt{s}/\Lambda^{(3)}_{\overline{MS}}=9$.\\

\noindent {\bf Fig.~15} Same as in Fig.~9, but for $n_{f}=3$ and
 $\sqrt{s}/\Lambda^{(3)}_{\overline{MS}}=9$. Only the $l=2$
 region has been indicated.\\

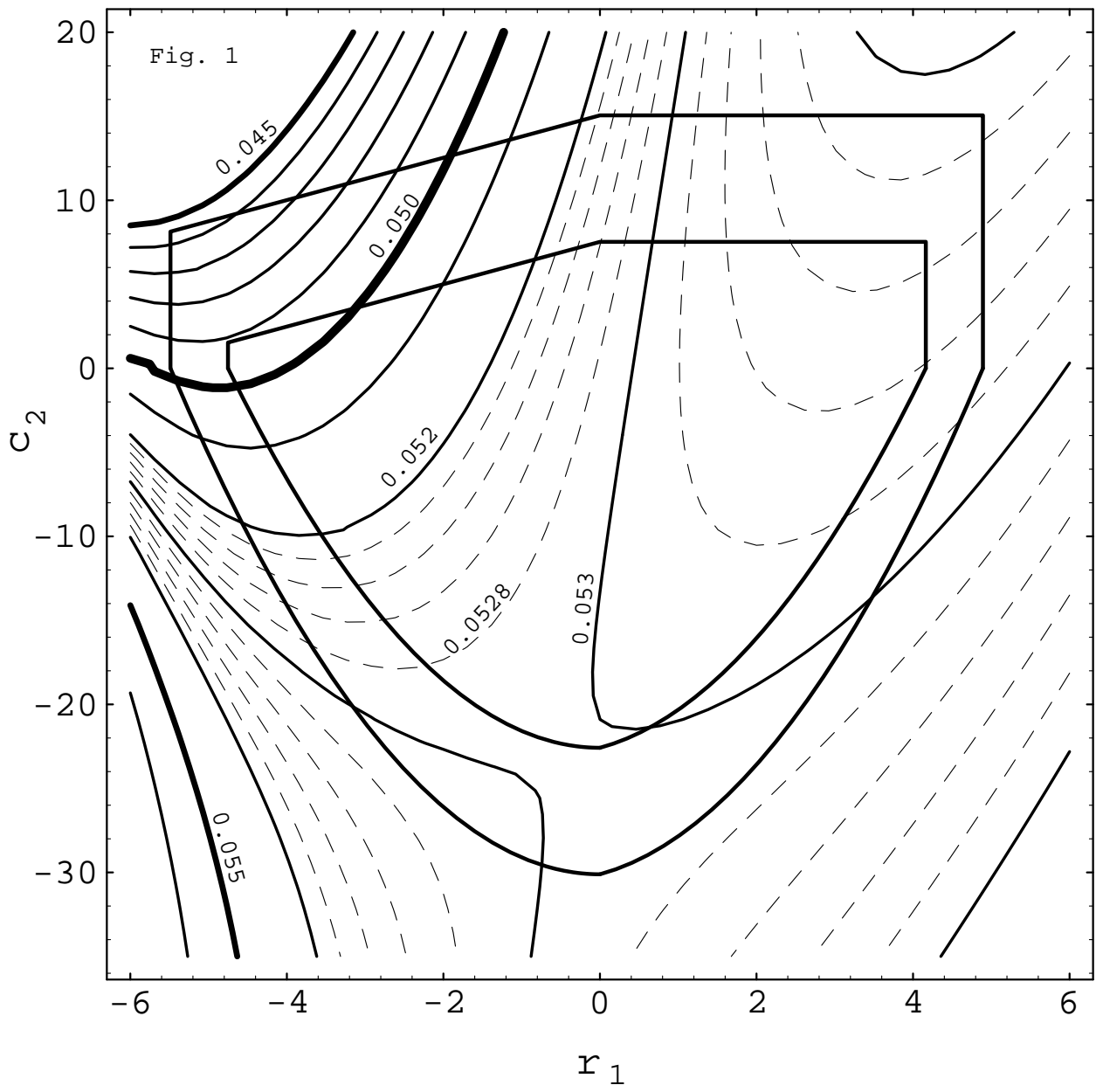
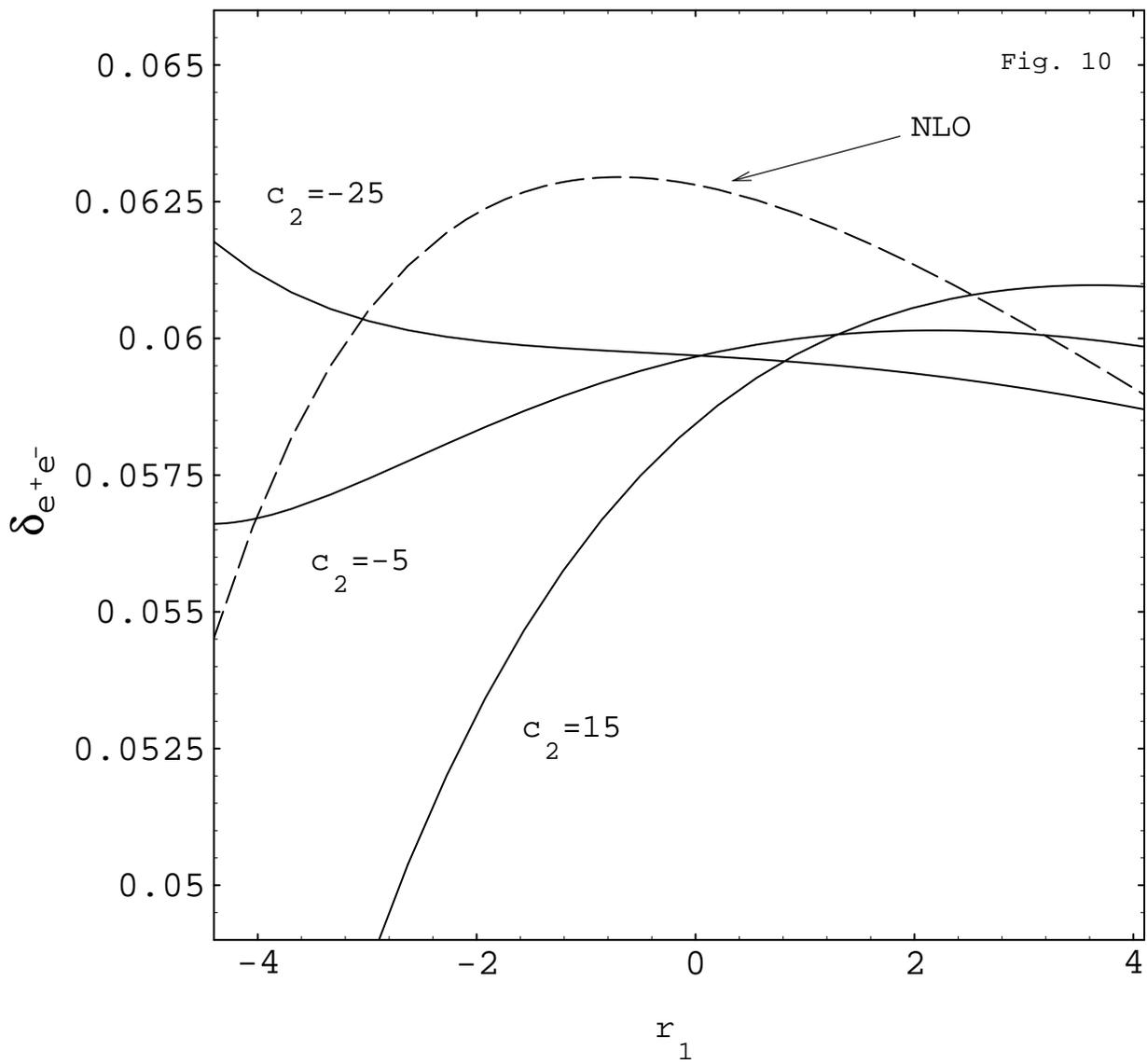
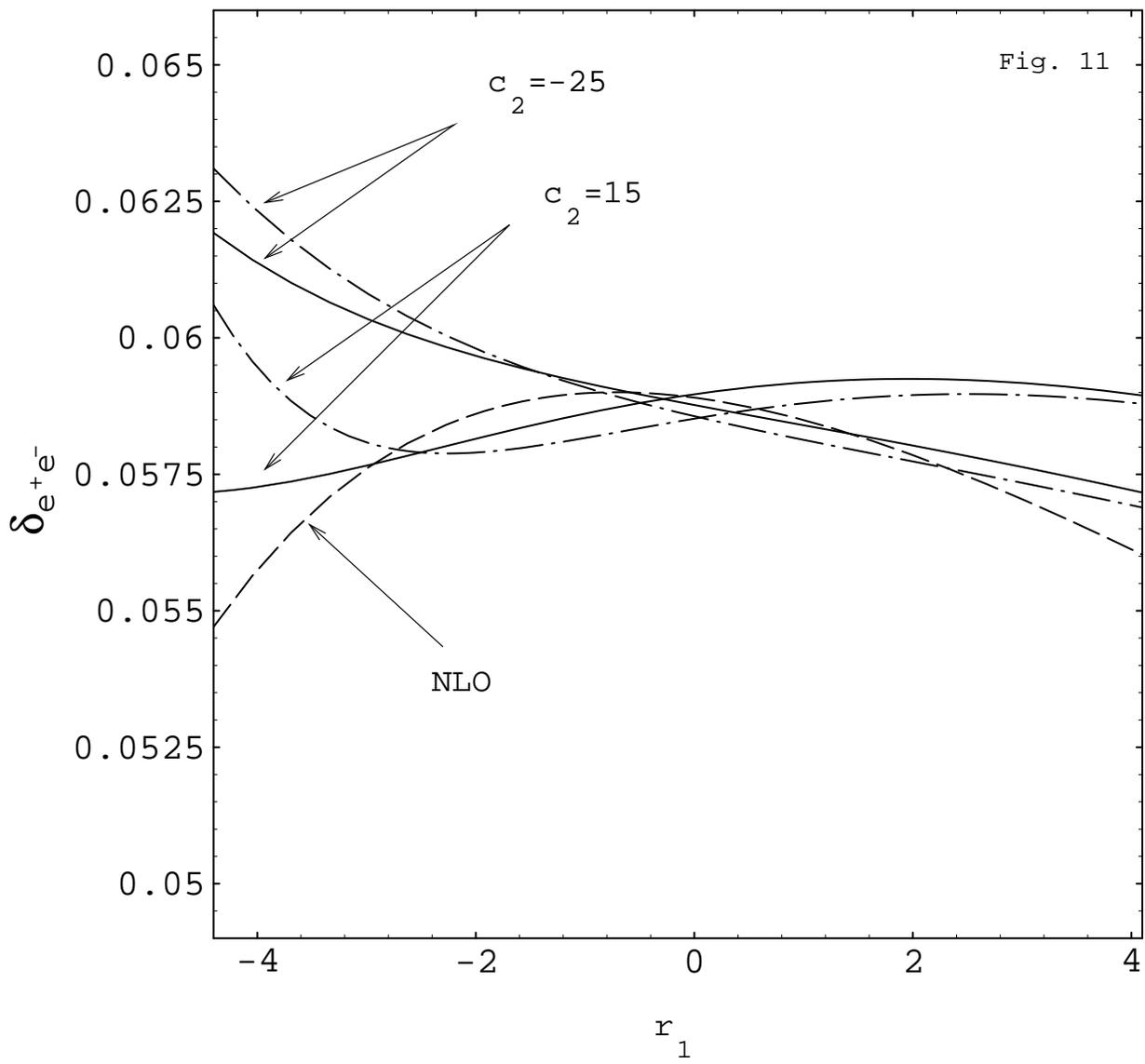
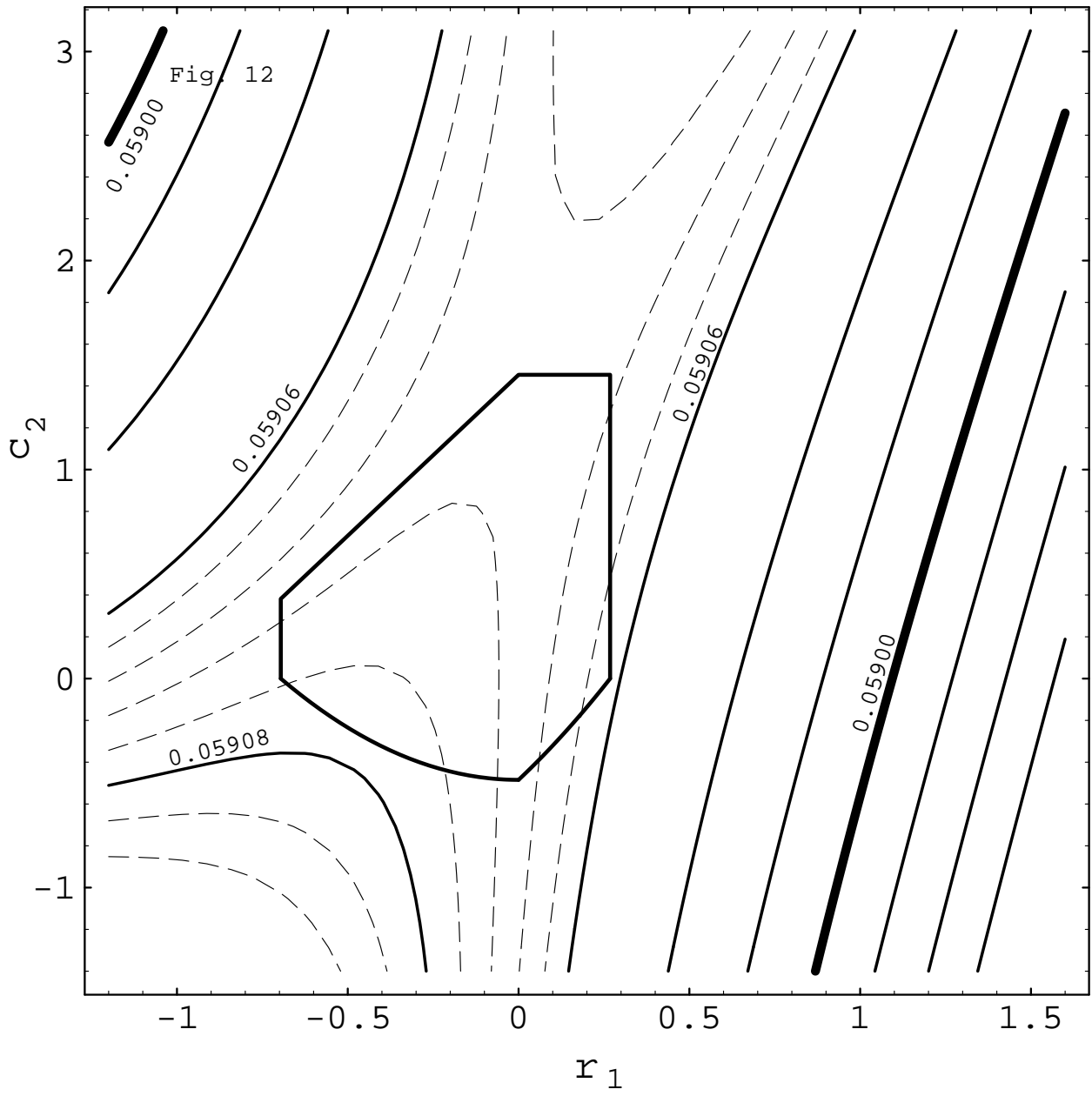
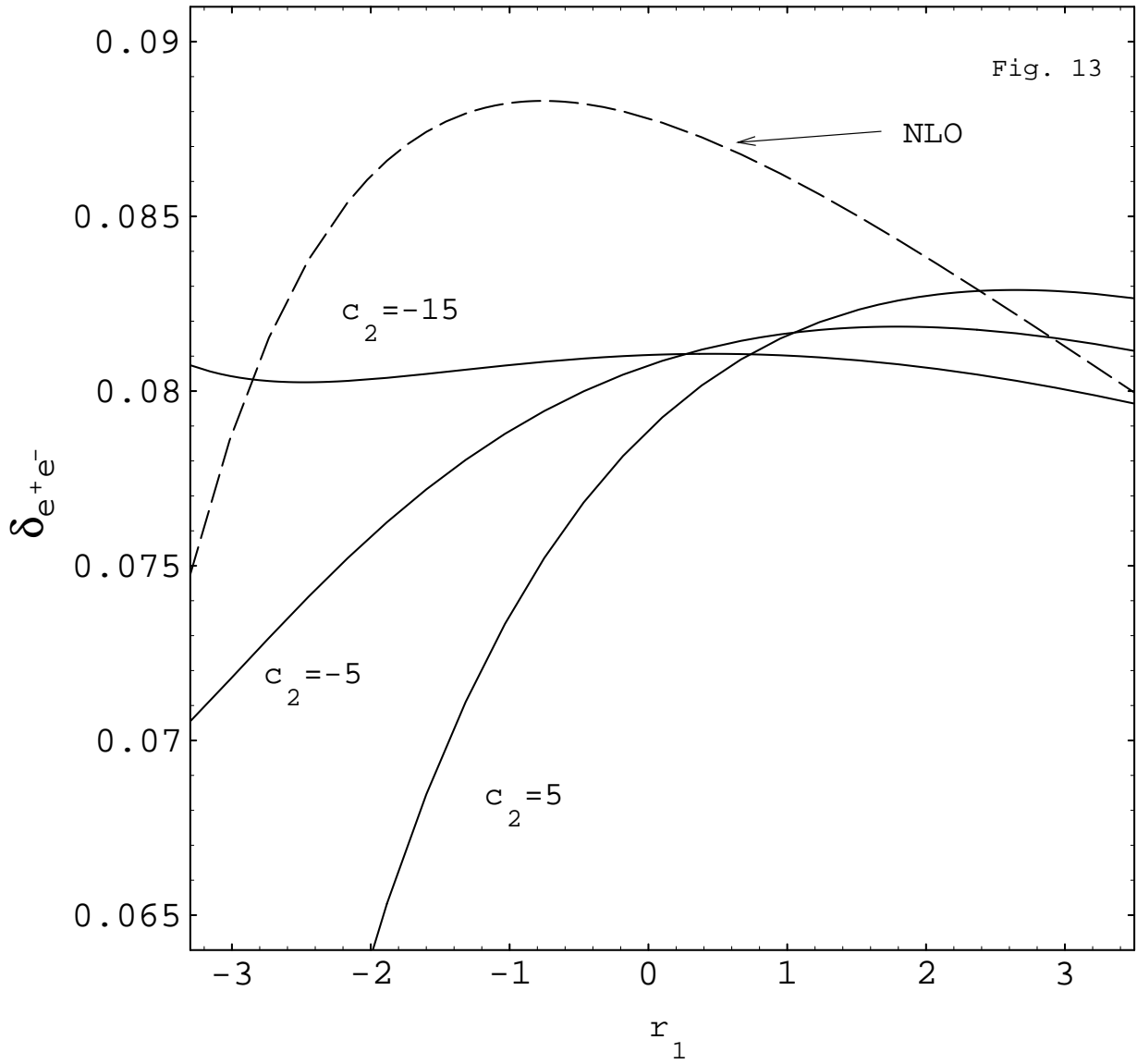
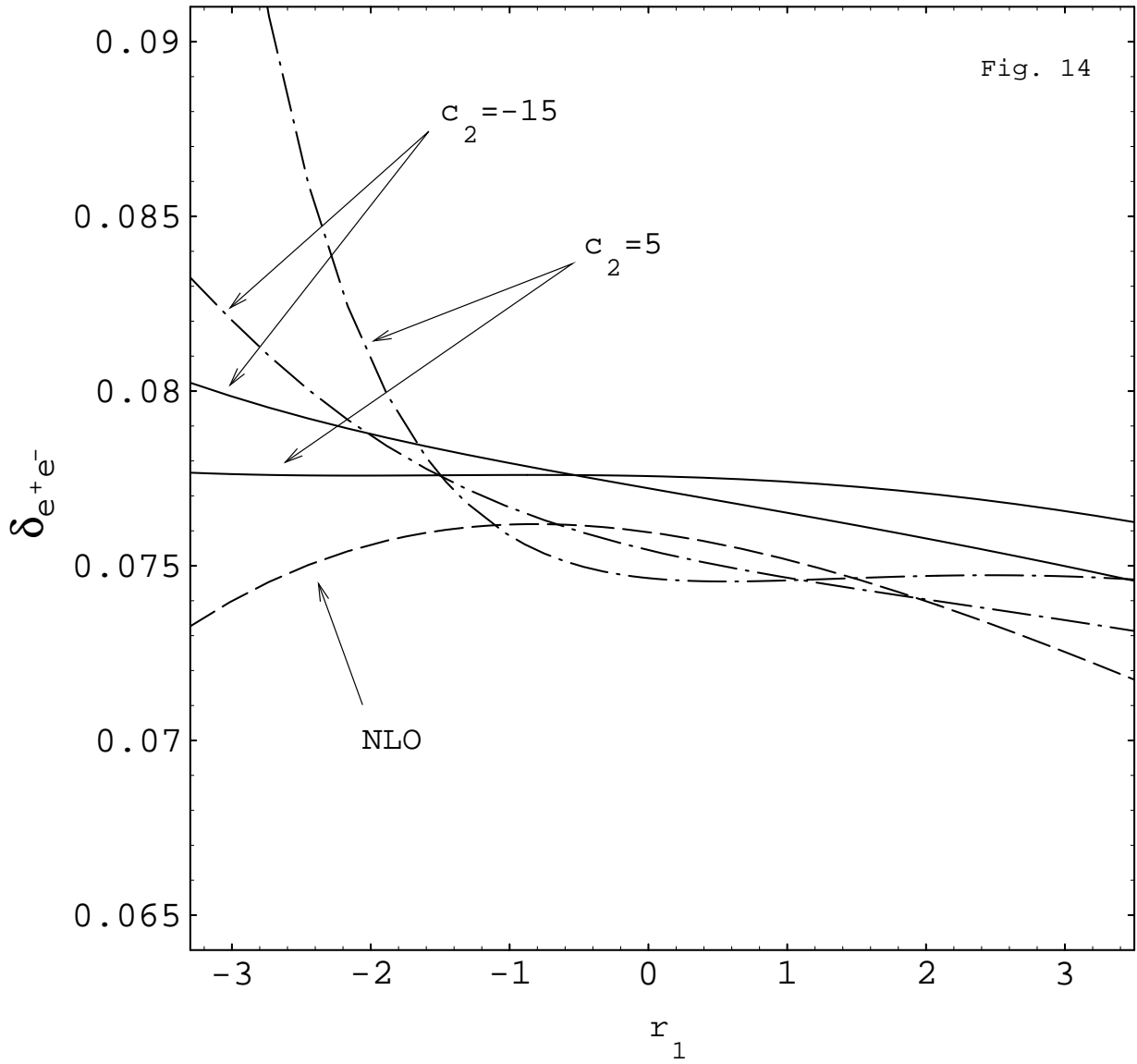
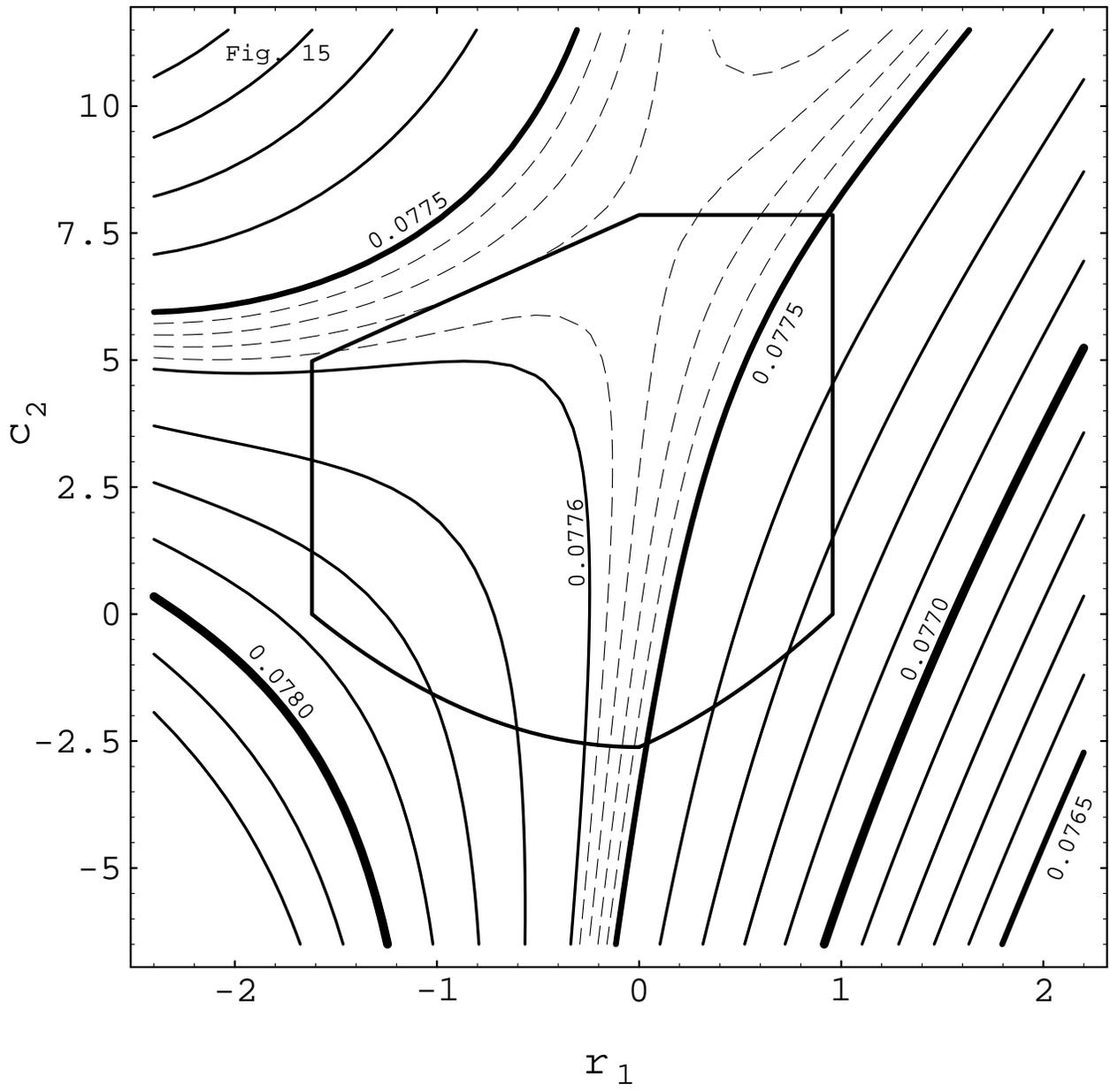
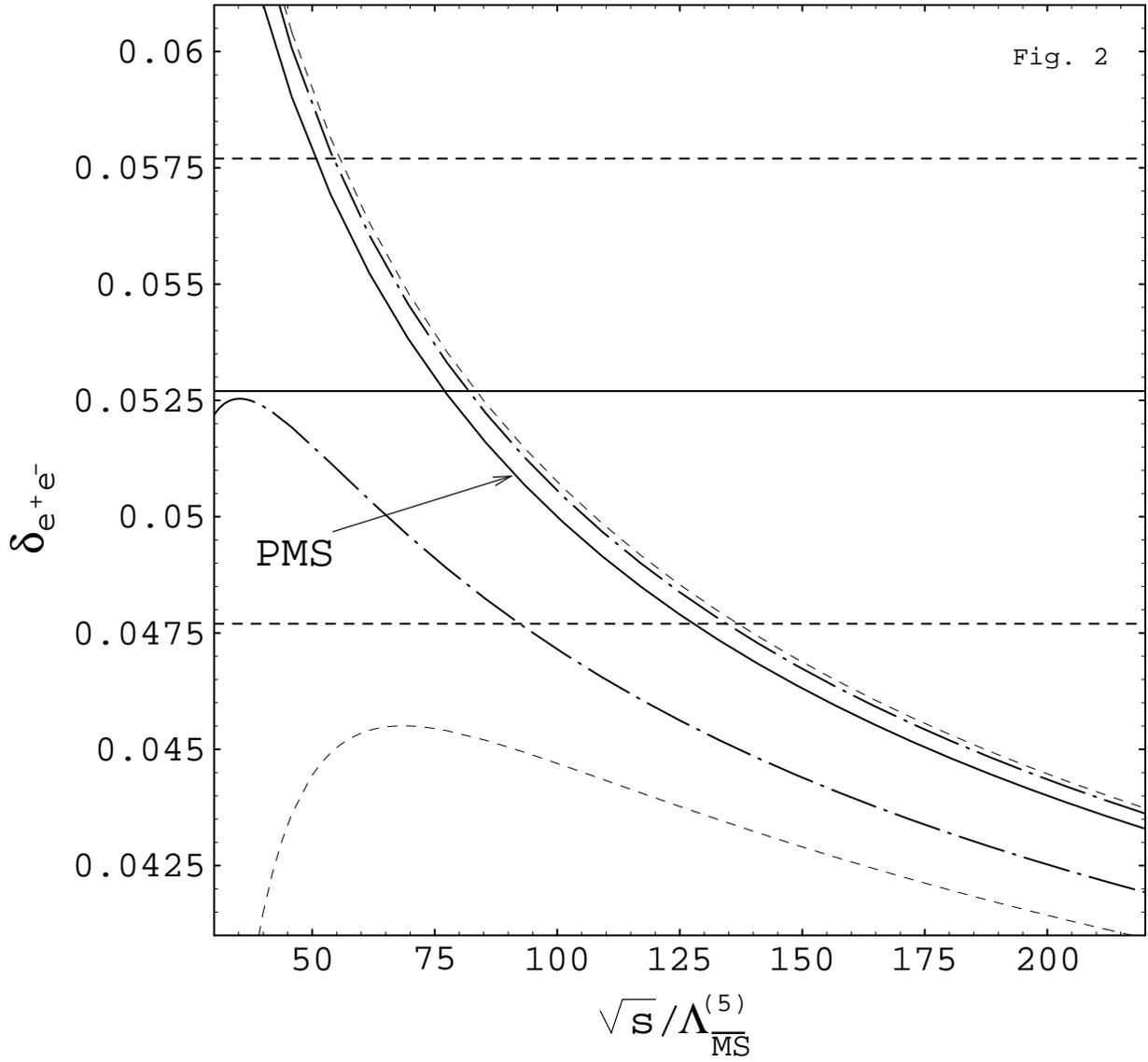
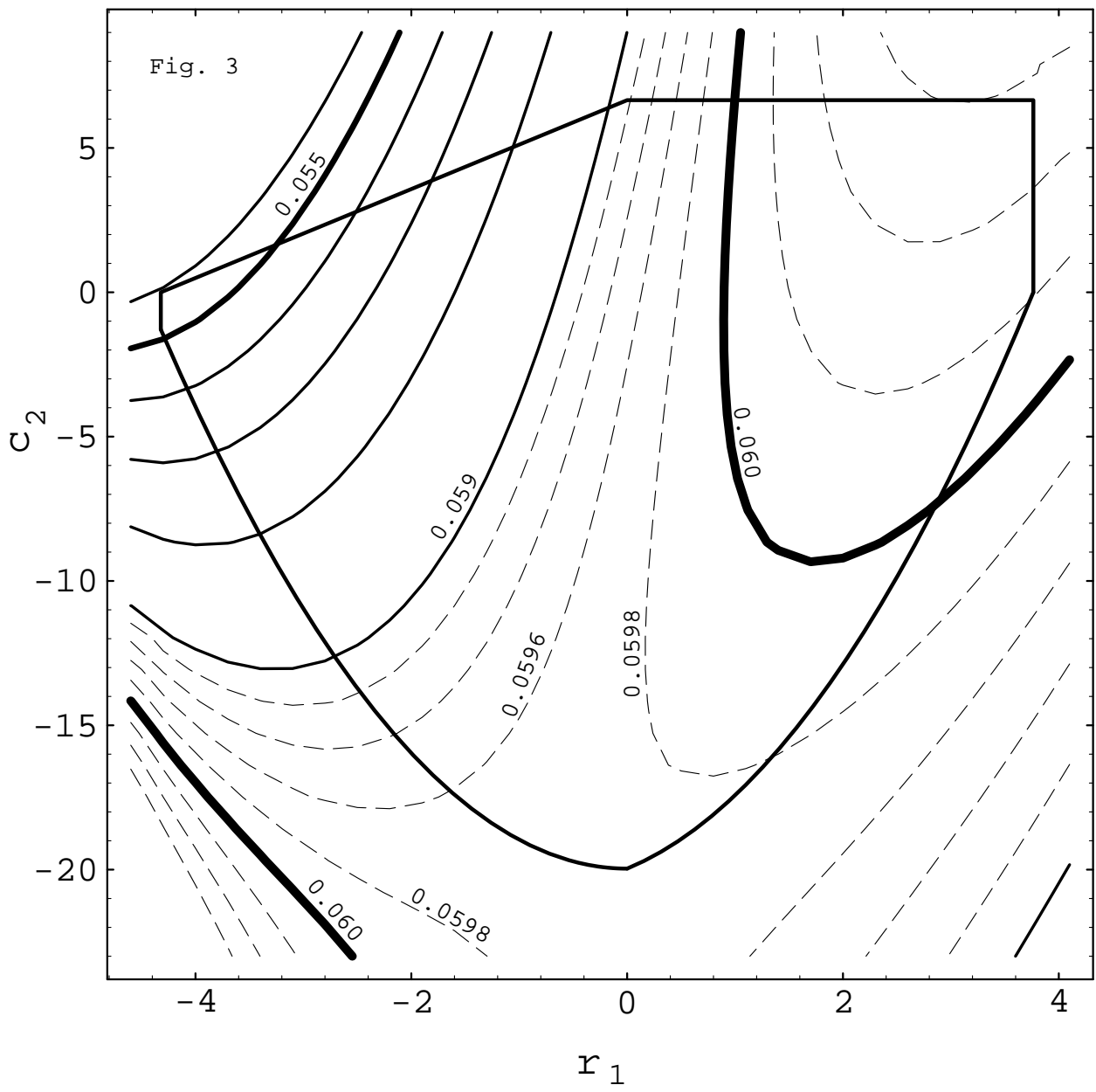
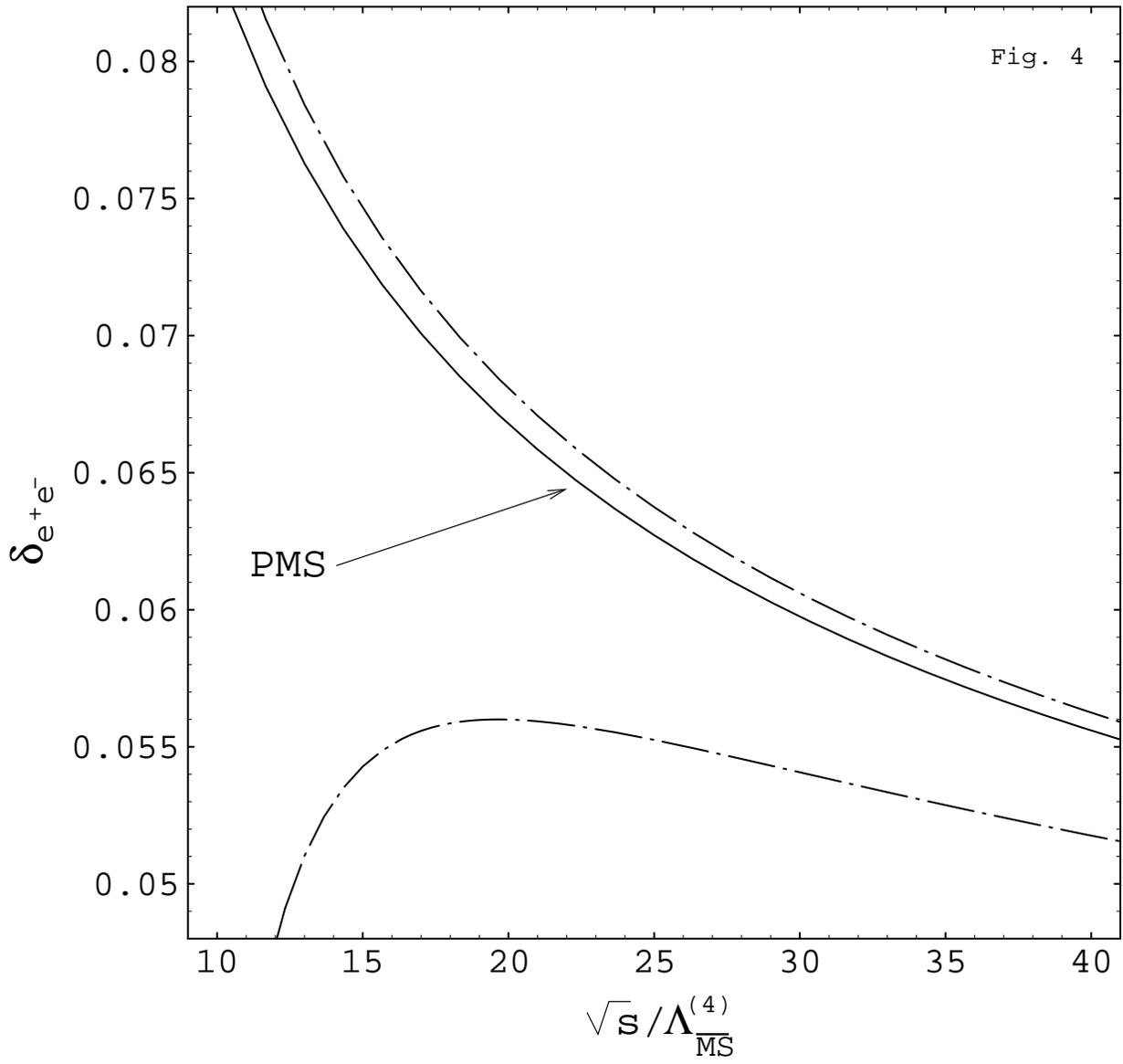
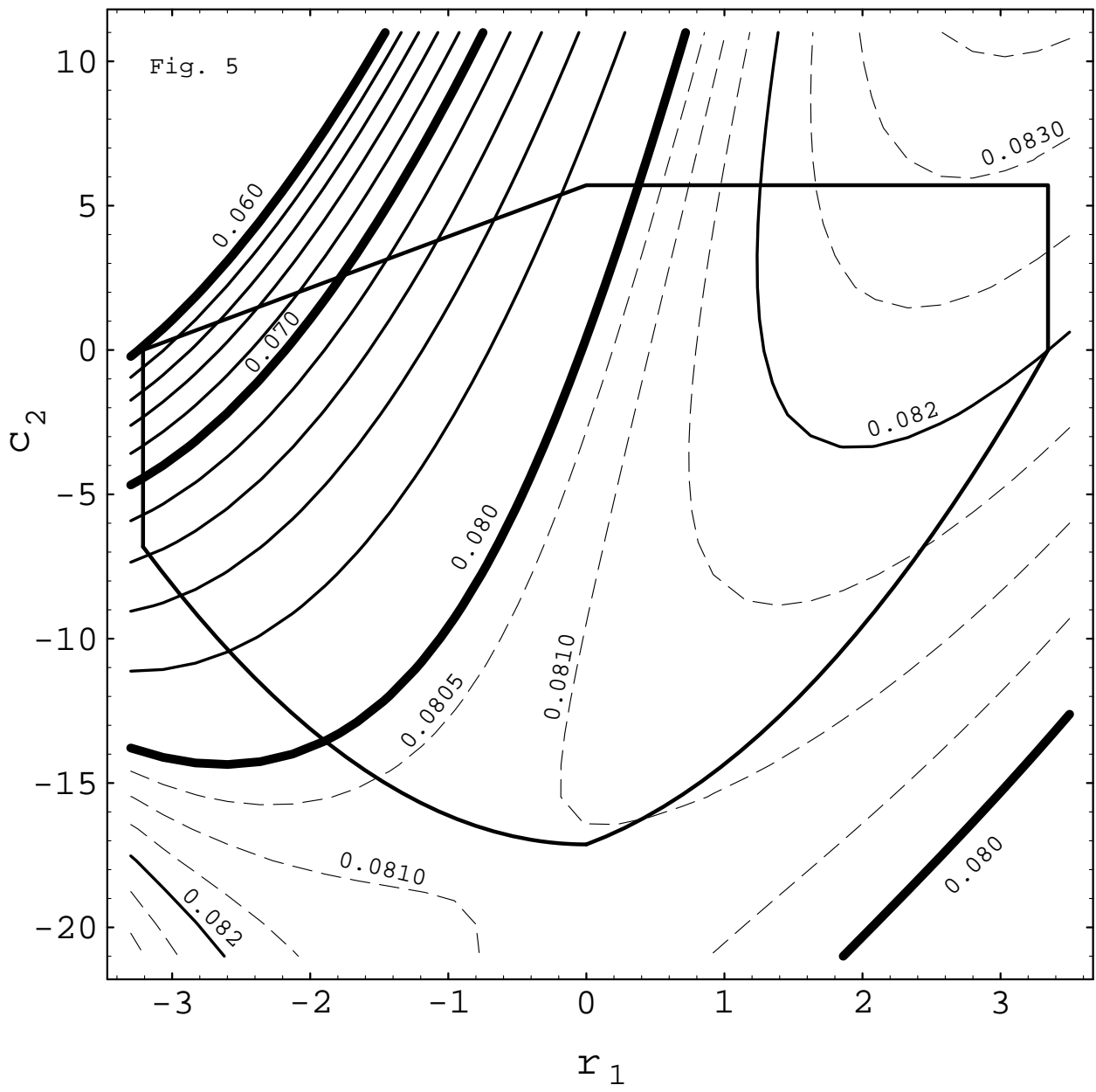
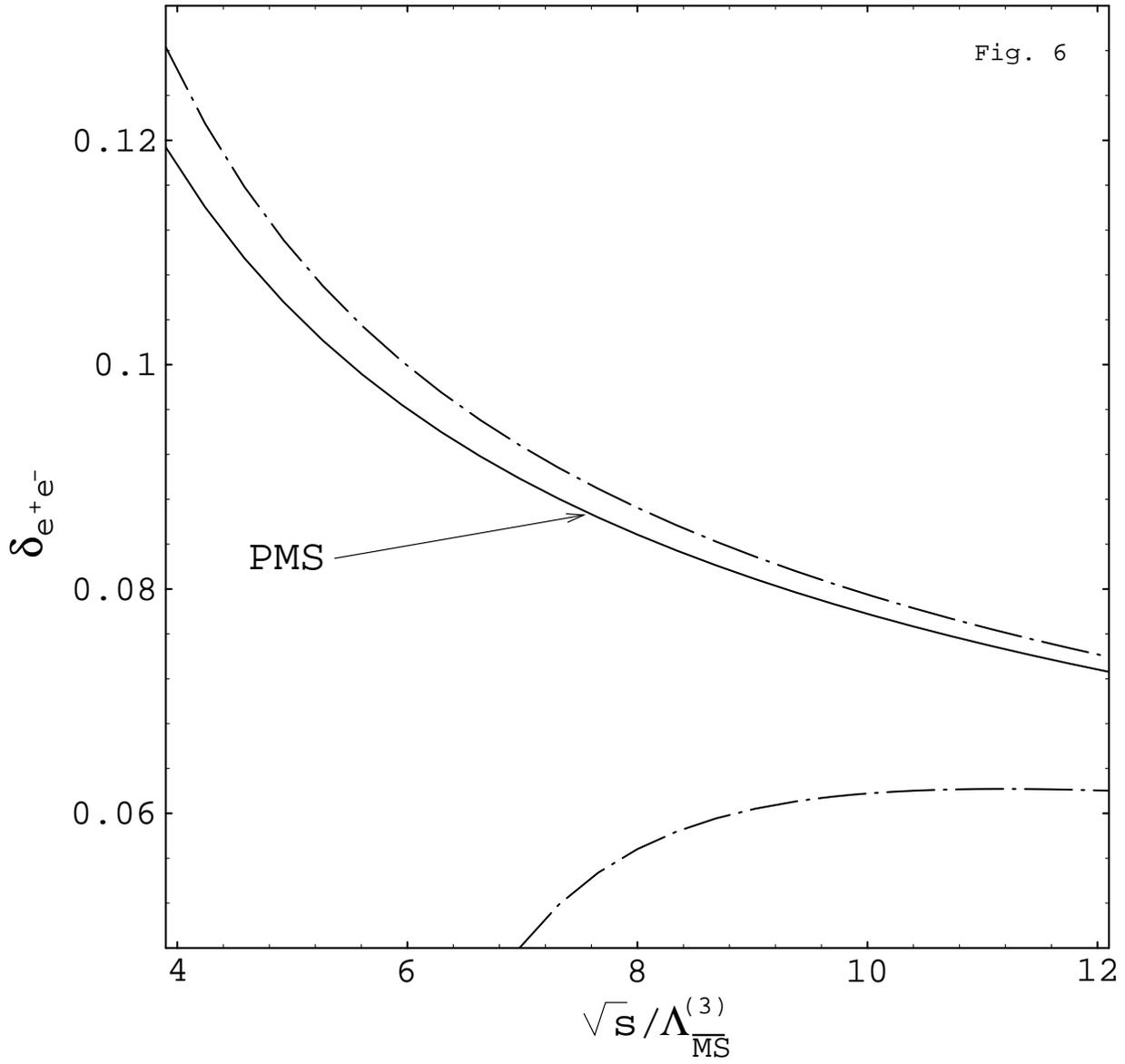
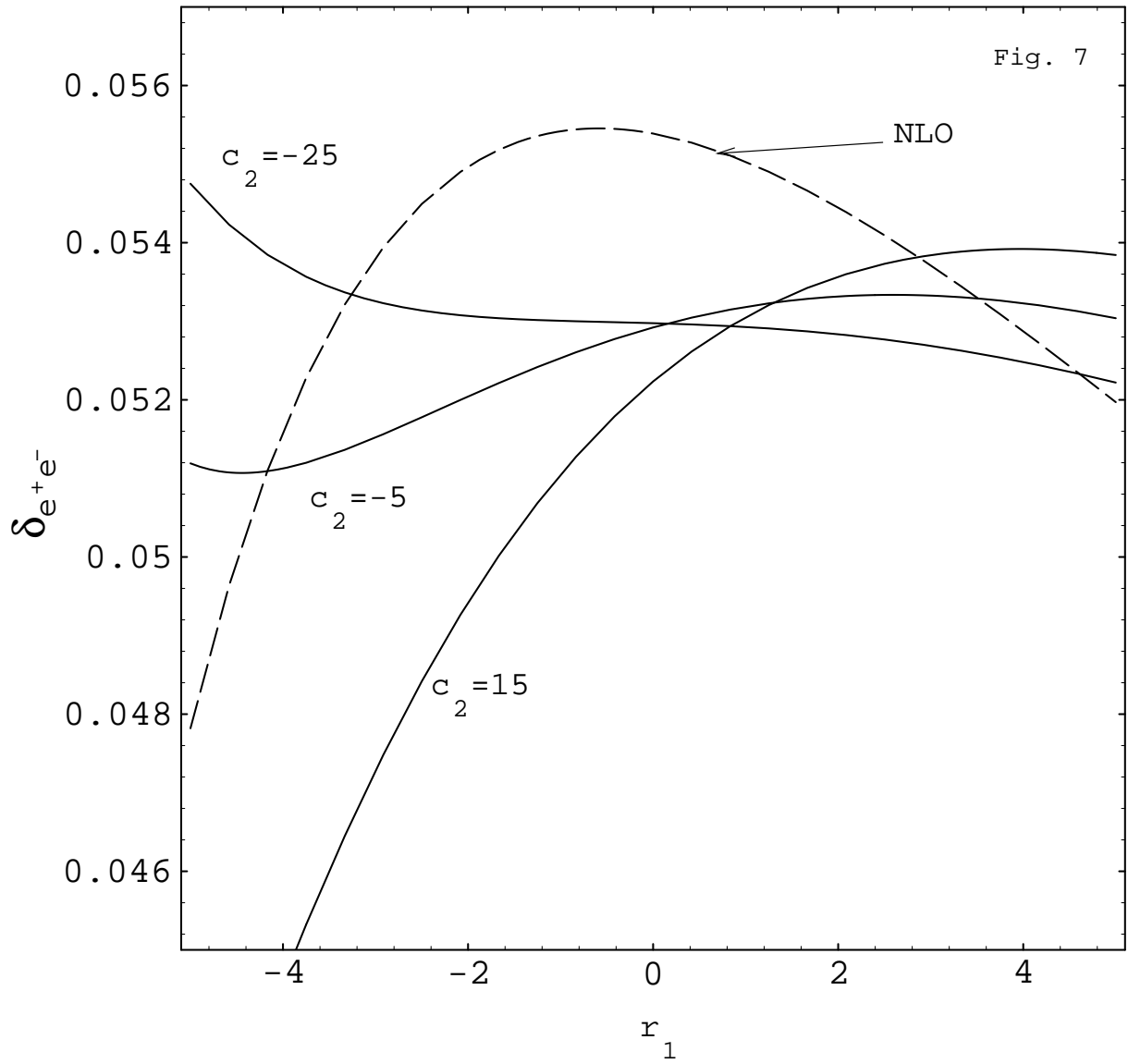
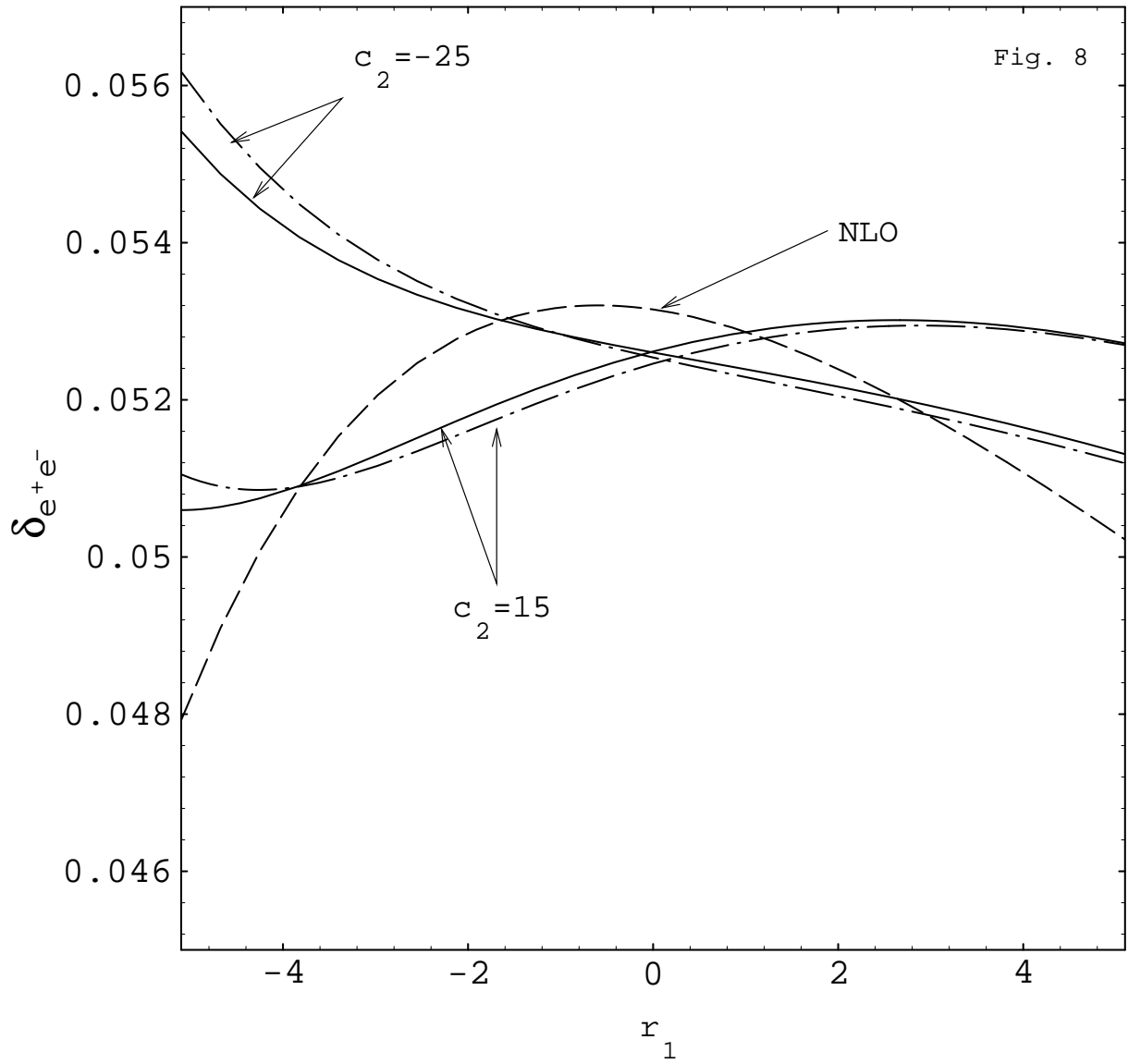
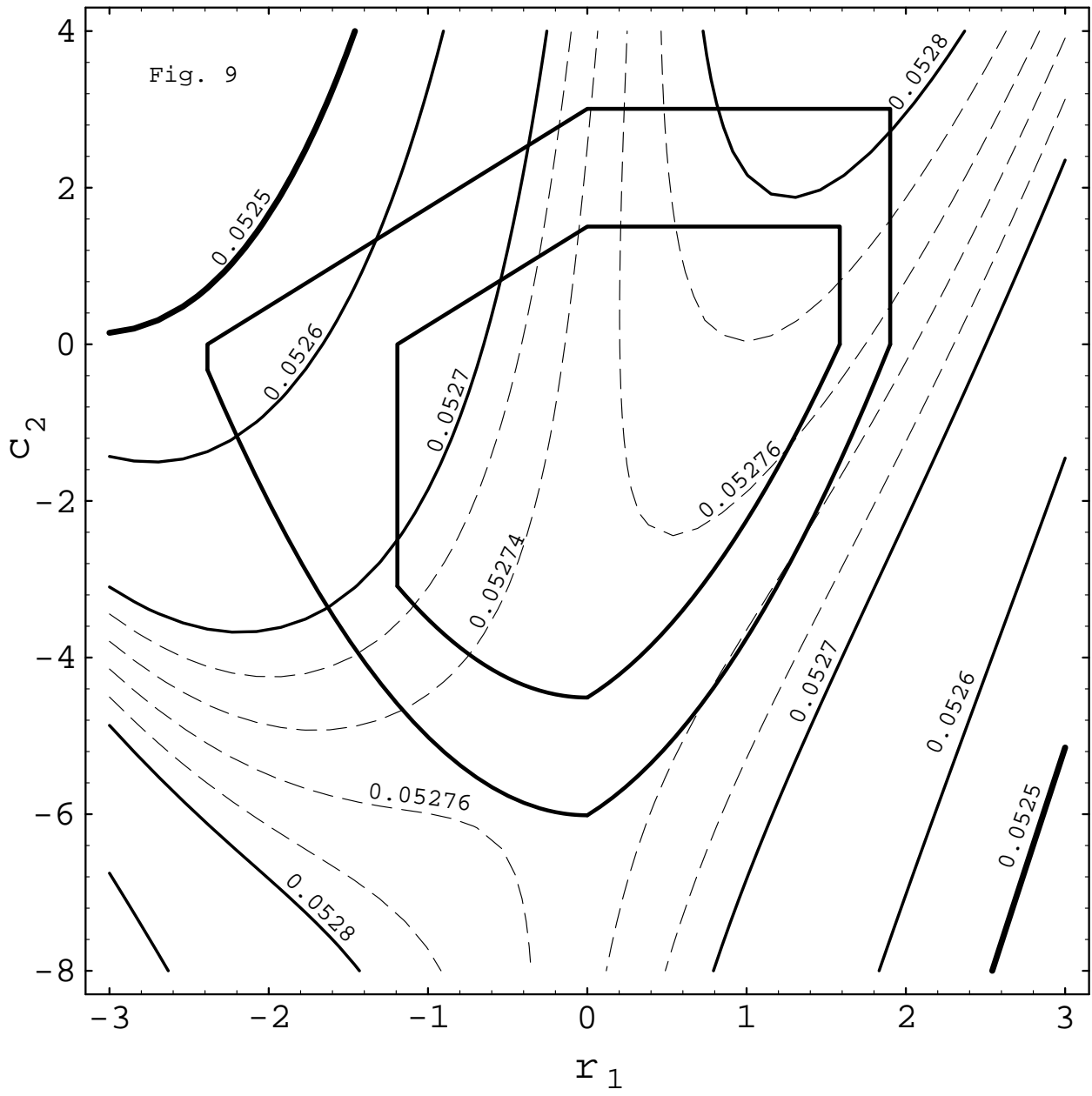
\end{document}